# Transformation and amplification of light modulated by a traveling wave with a relatively low frequency

M. Sumetsky

Aston Institute of Photonic Technologies, Aston University, Birmingham B4 7ET, UK
*e-mail*: m.sumetsky@aston.ac.uk

The behavior of electromagnetic waves in a medium modulated in time and space, largely investigated decades ago, has recently attracted renewed interest. Here, we address an intriguing problem of this research: can light with an initial frequency $\omega_0$ be amplified solely pumped by a traveling wave with a much lower frequency $\omega_p \ll \omega_0$? Generally, the bandwidth of the modulation-induced optical frequency comb spectrum can be substantially broadened when the phase velocity of the traveling wave, $v_p$, approaches the phase velocity of light, $v_0$. However, in realistic photonic waveguides, the amplification effect remains small due to the unfeasible modulation and waveguide parameters required. In contrast, we demonstrate that modulating an optical resonator by a traveling wave having the frequency $\omega_p$ and phase velocity $v_p$ much smaller than the frequency $\omega_0$ and phase velocity $v_0$ of light can result in large light amplification accompanied by conversion to multiple comb lines within a relatively small frequency band.

## I. INTRODUCTION

The growing interest in exploring wave propagation through media *parametrically modulated in time and space* is driven by its intriguing features – not possible in the stationary case – along with current and potential applications [1-3]. Numerous earlier and recent papers investigated modulation-induced amplification of waves [4-18], signal processing [19-24], frequency comb generation [25-28], and sideband transitions, including the effects of propagation nonreciprocity and complete inelastic transparency [29-39]. Temporal modulation can also create dynamic bandgaps where waves with certain frequencies are trapped in localized regions [35, 40, 41]. Modulating the properties of a medium can affect the group velocity of waves, resulting in slow or fast light [42]. Temporal modulation also allows for real-time wavefront control, enabling dynamic beam steering, focusing, and diffraction pattern manipulation, which are important in adaptive optics and beamforming technologies [43].

A crucial feature of wave propagation in a time-modulated medium is the potential for amplification. Modulation can transfer energy to the wave, enhancing its amplitude, or can extract energy from it, leading to attenuation. For example, the temporal modulation of the medium refractive index $\Delta n(t) = \Delta n_p \cos(\omega_p t)$ with frequency $\omega_p$ close to a multiple of the input electromagnetic wave half-frequency, $\omega_0/2$, can lead to amplification or attenuation of this wave described by Floquet theory ( see e.g., [11, 44]). For applications in optics, the amplification is customary achieved by pumping with a high-power light whose frequency $\omega_p$ is *comparable* to the frequency of input light to be amplified, $\omega_p/\omega_0 \sim 1$ [45]. For example, in Brillouin and Raman lasers, the acoustic and molecular vibrations are excited by a pump light with a frequency $\omega_p$ that is relatively close to the frequency of amplified light, commonly with $|\omega_p - \omega_0| \ll \omega_p$ [46-49].

However, is it possible to amplify an optical wave with a frequency $\omega_0$ in a realistic photonic circuit modulated *solely by a travelling wave with a much smaller frequency* $\omega_p \ll \omega_0$ (e.g., by an acoustic or RF wave) in the absence of pumping light? For an ideal waveguide with *negligible dispersion and losses over a large bandwidth* $\Delta\omega_B \gtrsim \omega_0$, a positive answer to this question was given several decades ago [8, 9]. The authors of Ref. [8, 9] (and, independently, the authors of Ref. [14]) found the exact solution of this problem for a one-dimensional propagation of a wave in a medium with constant impedance modulated by a traveling wave [8] and its asymptotic (eikonal, WKB) solution for a medium with constant permeability [9, 14]. It was shown that, under these conditions, amplification is indeed possible if the phase velocity of light $v_0$ is close to the phase velocity $v_p$ of the traveling wave. These results are irrelevant to realistic photonic circuits since their transmission loss and dispersion are *never negligible within the frequency bandwidth* $\Delta\omega_B \gtrsim \omega_0$ required for the observation of substantial amplification [6, 9, 15]. Consequently, the intriguing question of whether light amplification can be achieved by modulating a photonic circuit solely with a traveling wave having frequency $\omega_p \ll \omega_0$ remains open.

Here, we suggest an answer to this question. First, we explore the described problem using the eikonal approximation, which is valid when the modulation is slow in time and space, i.e., when it has a relatively small frequency $\omega_p \ll \omega_0$ (assumed throughout the paper), and wavenumber, $k_p = \frac{\omega_p}{v_p} \ll k_0 = \frac{\omega_0}{v_0}$ [50-54]. We show that, due to the requirement of the broadband and lossless transmission [8, 9], the substantial amplification of light by a low frequency modulation is unfeasible in realistic optical waveguides if the phase velocities of the modulating wave and light have the same order, $v_p \sim v_0$, and, in particular, are equal to each other, $v_p = v_0$. For a very small phase velocity $v_p$, when the wave numbers of light and modulating wave become comparable, $\frac{\omega_p}{v_p} \sim \frac{\omega_0}{v_0}$, the eikonal approximation fails, but a regular perturbation theory over the modulation amplitude of refractive index comes into force. Using this theory, we show that the amplification effect can significantly increase with a decrease in the phase velocity of the modulating wave, $v_p$. However, it still remains small for realistic waveguides and feasible modulation methods. In contrast, we demonstrate that light propagating forward through an optical racetrack lithium niobate resonator with potentially feasible characteristics can be significantly amplified by an acoustic wave. The developed theory is valid under the assumption of sufficiently small power of light inside the resonator introducing negligible nonlinear effects.

## II. TRANSFORMATION AND AMPLIFICATION OF LIGHT IN AN OPTICAL WAVEGUIDE IN THE EIKONAL APPROXIMATION

In this section, we consider the propagation of an optical wave along a dispersionless waveguide modulated by a low-frequency traveling wave, illustrated in Fig. 1 in the eikonal approximation. We assume that the input wave has the initial phase velocity $v_0$ and frequency $\omega_0$, while the modulating traveling wave has the phase velocity $v_p$ and a much smaller frequency $\omega_p \ll \omega_0$. The one-dimensional wave propagation is described by the wave equation

$$\left(n^2 E\right)_{tt} - c^2 E_{xx} = 0\,, \tag{1}$$

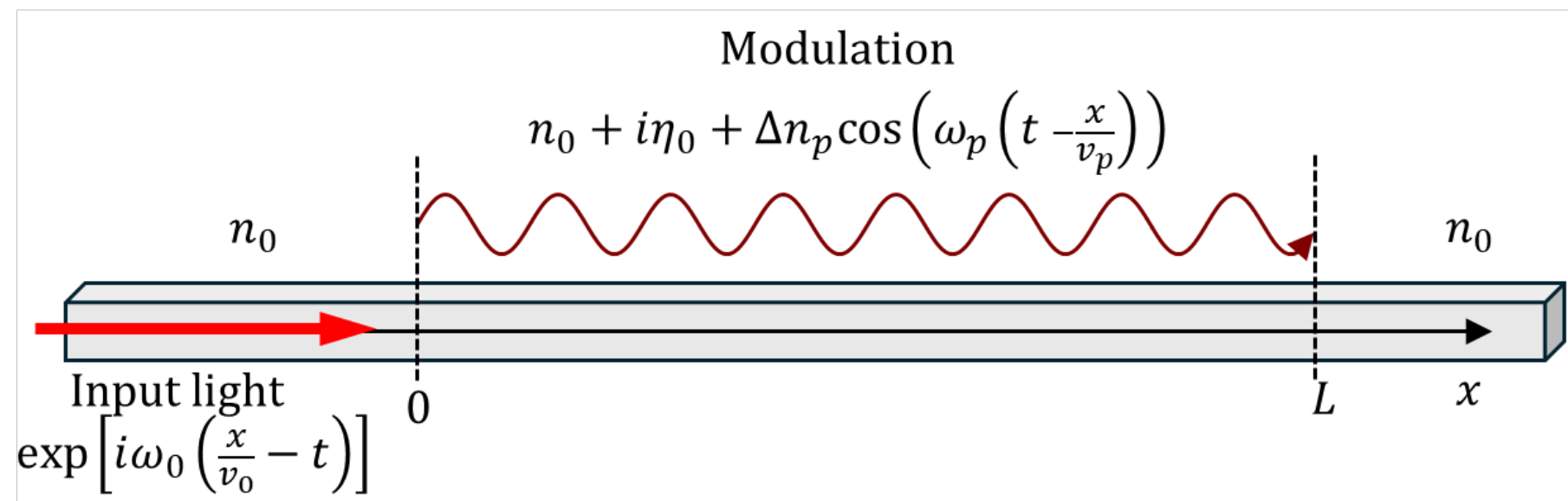

**Fig. 1.** An optical waveguide with the refractive index modulated by a traveling wave along the interval $(0, L)$.

where the subindices denote partial derivatives, and $c$ is the speed of light. The dependence of refractive index on time $t$ and coordinate $x$ along the waveguide is set to

$$\begin{gathered} n(x,t) = n_0 + \Delta n(x,t), \\ \Delta n(x,t) = \begin{cases} i\eta_0 + \Delta n_p \cos\left(\omega_p\left(t - \dfrac{x}{v_p}\right)\right) & 0 < x < L \\ 0 & \text{elsewhere} \end{cases} \end{gathered} \tag{2}$$

The waveguide propagation loss $\alpha_0$ is expressed through the imaginary part of the refractive index $\eta_0$ introduced in this equation as:

$$\alpha_0 = \frac{\eta_0 \omega_0}{c} \tag{3}$$

The boundary condition for the solution of Eq. (1) along a waveguide (Fig. 1) is defined by the input wave with frequency $\omega_0$ and phase velocity $v_0 = c/n_0$:

$$E^{(in)}(x,t) = \exp\left[i\omega_0\left(\frac{x}{v_0} - t\right)\right], \quad v_0 = \frac{c}{n_0}, \quad x < 0. \tag{4}$$

We notice that even for a small modulation amplitude $\Delta n_p \ll n_0$, solution of Eq. (1) by the perturbation theory over $\Delta n_p$ is incorrect if the phase velocities $v_p$ and $v_0$ are close to each other so that $|v_p - v_0|/v_0 \sim \Delta n_p/n_0 \ll 1$. The problem one faces here is similar to the small denominator problem (see, e.g., [55]). Alternatively, the propagation of waves in a medium, whose parameters are slowly varying in space and time, can be described in the eikonal approximation also known as the geometric optics approximation in electromagnetic theory [8, 9, 50, 51] and the WKB and semiclassical approximation in quantum mechanics [52, 53].

### A. Solution of the wave equation in the eikonal (WKB) approximation

Application of the eikonal approximation requires that the parameters of the optical waveguide vary slowly both in time and space. Specifically, the frequency $\omega_p$ and wavenumber $k_p$ of the traveling wave should be small compared to the frequency $\omega_0$ and wavenumber $k_0$ of the input wave:

$$\begin{aligned} &\omega_p \ll \omega_0, \quad k_p \ll k_0, \\ &k_0 = \frac{\omega_0}{v_0}, \quad k_p = \frac{\omega_p}{v_p}. \end{aligned} \tag{5}$$

We also assume that the material waveguide loss $\eta_0$ is relatively small, $\eta_0 \ll n_0$. Then, the solution of Eq. (1) $E(x,t)$ in the region $0 < x < L$ can be found by the eikonal (semiclassical) theory [50, 51]. In this theory, the solution of Eq. (1) is presented as $E(x,t) = \exp\,(iS_0(x,t)/\varepsilon)\sum_{n=0}^{\infty}\varepsilon^n U_n(x,t)$ where the small parameter $\varepsilon = \max(\omega_p/\omega_0,\, k_p/k_0) \ll 1$ and $S(x,t) = S_0(x,t)/\varepsilon$ is the eikonal satisfying the equation $n^2(x,t)S_t^2 - c^2 S_x^2 = 0$. For the small modulation, $\Delta n_0 \ll n_0$, or, alternatively, for the modulation adiabatically switching on near the coordinate $x = 0$ and off near $x = L$ (the switching region is not illustrated in Fig. 1), we can ignore the reflected waves at $x = 0$ and $x = L$. Then, similar to calculations of Ref. [8], we find the zero-order in $\varepsilon$ asymptotic solution of Eq. (1) with the refractive index defined by Eq. (2) and the boundary condition of Eq. (4) (see Appendix A):

$$\begin{aligned} &E(x,t) = U_0(x,t)\exp\big(iS(x,t)\big), \\ &S(x,t) = -\omega_0 \overline{t}\,(\xi(x,t)), \\ &U_0(x,t) = \frac{\Big(1+\mu\cos\big(\omega_p \overline{t}\,(\xi(x,t))\big)\Big)\sqrt{1+\mu\delta\cos\big(\omega_p \overline{t}\,(\xi(x,t))\big)}}{\left(1+\mu\cos\left(\omega_p\left(t-\dfrac{x}{v_p}\right)\right)\right)\sqrt{1+\mu\delta\cos\left(\omega_p\left(t-\dfrac{x}{v_p}\right)\right)}}, \end{aligned} \tag{6}$$

$$\overline{t}\,(\xi) = -\frac{2}{\omega_p}\,\Xi\left(\sqrt{\frac{1+\mu}{1-\mu}},\, \frac{\sqrt{1-\mu^2}}{2v_0 v_p}\big(v_p - v_0\big)\omega_p \xi\right), \tag{7}$$

$$\xi(x,t) = x - \frac{2v_0 v_p}{\omega_p\big(v_p - v_0\big)\sqrt{1-\mu^2}}\,\Xi\left(\sqrt{\frac{1-\mu}{1+\mu}},\, \frac{\omega_p}{2}\left(t-\frac{x}{v_p}\right)\right), \tag{8}$$

$$\mu = \frac{\Delta n_p}{n_0 \delta}, \quad \delta = \left(1 - \frac{v_0}{v_p} + i\frac{\eta_0}{n_0}\right). \tag{9}$$

Here, function $\Xi(x,y)$ is the smooth continuation of $\arctan\,(x \cdot \tan(y))$ as a function of $y$ (see Eq. (A12)

in Appendix A).

It follows from Eqs. (6)-(9) that the field $E(x,t)$ is periodic in time with the period $2\pi/\omega_p$ though may be aperiodic in space. In the major calculations below, we ignore the factors under the square roots in Eq. (6) since we always assume that $\mu\delta = \Delta n_p/n_0 \ll 1$. The behavior of solution $E(x,t)$ at position $x$ is characterized by the complex-valued (in the presence of losses) *synchronization parameter* $\mu$ introduced in Eq. (9). For small losses considered below, $\eta_0 \ll \Delta n_p$, this parameter is the ratio of the relative modulation amplitude $\Delta n_p/n_0$ and the relative proximity of velocities of the input wave and the modulating traveling wave, $1 - v/v_p$. Following Ref. [8, 9], we call the modulation *asynchronous* if $|\mu| < 1$, call it *completely asynchronous* if $|\mu| \ll 1$, (in particular, call it *instantaneous* if $v_p = \infty$), call it *synchronous* if $|\mu| > 1$, and call it *completely synchronous* if $|\mu| \gg 1$ (in particular, if $v_p = v$). While the solution defined by Eqs. (6)-(9) is quasiperiodic in space in the case of asynchronous modulation, it can exponentially grow in space for synchronous modulation.

### B. The wave amplification effect in the absence of losses

Under the condition of negligible material losses, $\eta_0 = 0$, the synchronization parameter $\mu$ is real and there exist two qualitatively different cases of the spatially unstable ($|\mu| > 1$) and spatially stable ($|\mu| < 1$) solutions, corresponding to the synchronous and the asynchronous cases introduced above. Fig. 2 presents the characteristic behavior of the normalized field amplitude $|U_0(L,t)|$ at a fixed time $t = 0$ and time-averaged normalized wave power, an average of the squared amplitude $U_0(x,t)$ defined by Eq. (6) over the time period $2\pi/\omega_p$:

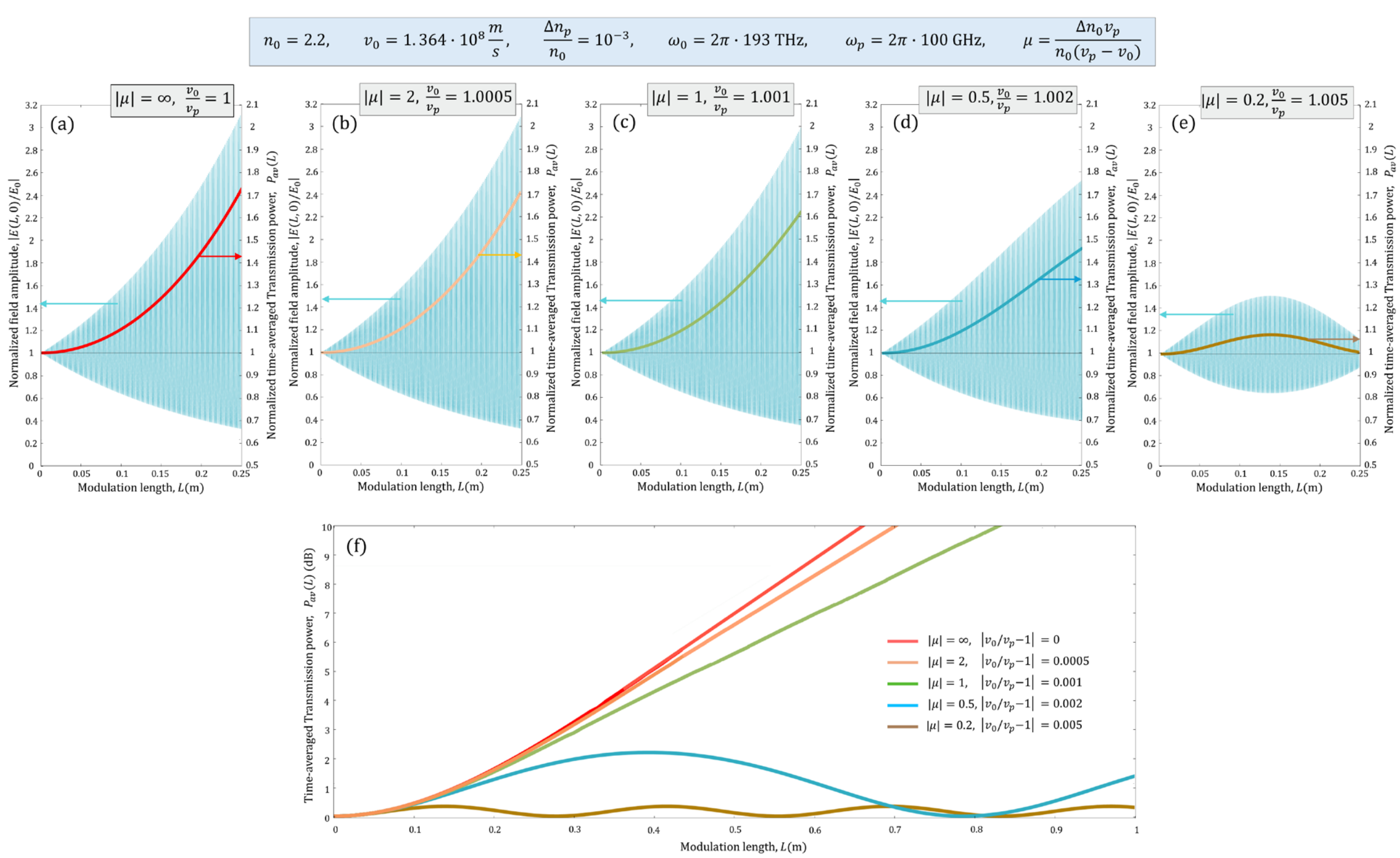

**Fig. 2.** The amplitude and time-averaged power of the output wave as a function of modulation length $L$ for different synchronization parameters $\mu$ corresponding to close phase velocities $v$ and $v_p$ of the input light and modulation. (a)-(e) The amplitude (left vertical axes) and normalized time-averaged power (right vertical axes) for modulation lengths $0 < L < 0.25$ m. (a) $|\mu| = \infty, v = v_p$ (completely synchronous case); (b) $|\mu| = 2, v = 1.001v_p$ (synchronous case); (c) $|\mu| = 1, v = 1.001v_p$; (d) $|\mu| = 0.5, v = 1.002v_p$ (asynchronous case); (e) $|\mu| = 0.2, v = 1.005v_p$ (asynchronous case); (f) Time-averaged power $P_{av}(L)$ for modulation lengths $0 < L <$

1 m for the relations between $v$ and $v_p$ of plots (a)-(e).

$$P_{av}(L) = \frac{\omega_p}{2\pi} \int_0^{2\pi/\omega_p} U_0(L,t)^2 \, dt \,. \tag{10}$$

Fig. 2(f) presents $P_{av}(L)$ as a function of modulation length $L$ for different synchronization parameters $\mu$. In this figure, we consider the propagation of light along the lithium niobate waveguide with refractive index $n_0 = 2.2$ modulated with relative amplitude $\Delta n_p/n_0 = 10^{-3}$. The input light frequency and modulation frequency are set to $\omega_0 = 2\pi \cdot 193$ THz and $\omega_p = 2\pi \cdot 100$ GHz. These values and the value of refractive index of the lithium niobate waveguide $n_0 = 2.2$ corresponding to the phase velocity of light $v_0 = \frac{c}{n_0} = 1.364 \cdot 10^8$ m/s are assumed throughout the paper.

The purpose of so large modulation amplitude $\Delta n_p$ and frequency $\omega_p$ considered is to evaluate the *largest possible effects of modulation* including the largest possible amplification. To visually resolve the fine spatial oscillations of the wave amplitude, Figs. 2(a)-(e) show the behavior of $|U_0(L,0)|$ (blue frequently oscillating curves) and $P_{av}(L)$ (bold curves of different color) along the interval $0 < L < 0.25$ m, while Fig. 2(f) shows the behavior of $P_{av}(L)$ over a longer interval $0 < L < 1$ m. It is seen that for $|\mu| > 1$ the wave amplitude $|U_0(L,t)|$ oscillates and grows with $L$ while its time-averaged value grows exponentially. In contrast, for $|\mu| < 1$, both $|U_0(L,0)|$ and $P_{av}(L)$ remain, respectively, quasiperiodic and periodic as a function of modulation length $L$.

It is also seen from Fig. 2(f) that, for the parameters considered, the dependencies of the wave amplitude $|U_0(L,0)|$ and the time-averaged power $P_{av}(L)$ on $L$ are similar for modulation lengths $L < 0.1$ m. Consequently, in this interval, the proximity to the full synchronization condition $v_p = v$ does not enhance the wave amplification, which is always small at these modulation lengths.

### C. The effect of losses

Material losses can significantly modify the behavior of the propagating wave shown in Fig. 2. Here, we find the effect of relatively small though practically feasible material losses for an ideally dispersionless waveguide. We note that the light propagation loss $\alpha_0$ (Eq. (3)) of a lithium niobate waveguide can be as small as 0.2 dB/m [56, 57] corresponding to $\eta_0 \sim 10^{-8}$ and commonly has the order of 10 dB/m or greater [22, 23]. In Fig. 3 we consider the effect of broadband dispersionless material losses for the phase velocity relation $v_p = 1.0005 v_0$. In this figure, the blue curves show the normalized wave power $|U_0(L,0)|^2$ as a function of modulation length at a fixed time, $t = 0$, the black curves show this dependence for the unmodulated waveguide, $\Delta n_p = 0$, and the red curves are the dependencies of time-averaged wave power $P_{av}(L)$ on the modulation length $L$. We notice that the relation $v_p = 1.0005 v_0$ considered in Fig. 3 corresponds to the synchronization parameter $|\mu| = 2$ similar to that for the relation $v_0 = 1.0005 v_p$ considered in Fig. 2(b). Consequently, the behavior of the field power shown in Fig. 2(b) and the corresponding time-averaged field power (orange curve in Fig. 2(f)) is similar to those in Fig. 3(a) for $\eta_0 = 0$. We find that the effect of attenuation for $\eta_0 = 10^{-8}$ (Fig. 3(b)) is small for modulation lengths $L < 0.3$ m and grows for larger $L$. This effect is much stronger for $\eta_0 = 10^{-7}$ (Fig. 3(b)) and for $\eta_0 = 10^{-6}$ (Fig. 3(c)).

### D. Wave propagation and wave spectrum in the lossless completely synchronous case $v_p = v_0$

Of special interest is the ideal lossless and completely synchronous case when the phase velocities $v$ and $v_p$ are equal, $v_p = v_0$, illustrated in Fig. 2(a) and (f). For optical wave propagation, this case is also referred to as the luminal case [16, 17]. Then, the expressions for the solution phase (eikonal) $S(x,t)$ and normalized amplitude $U_0(x,t)$ are simplified (see Appendix B):

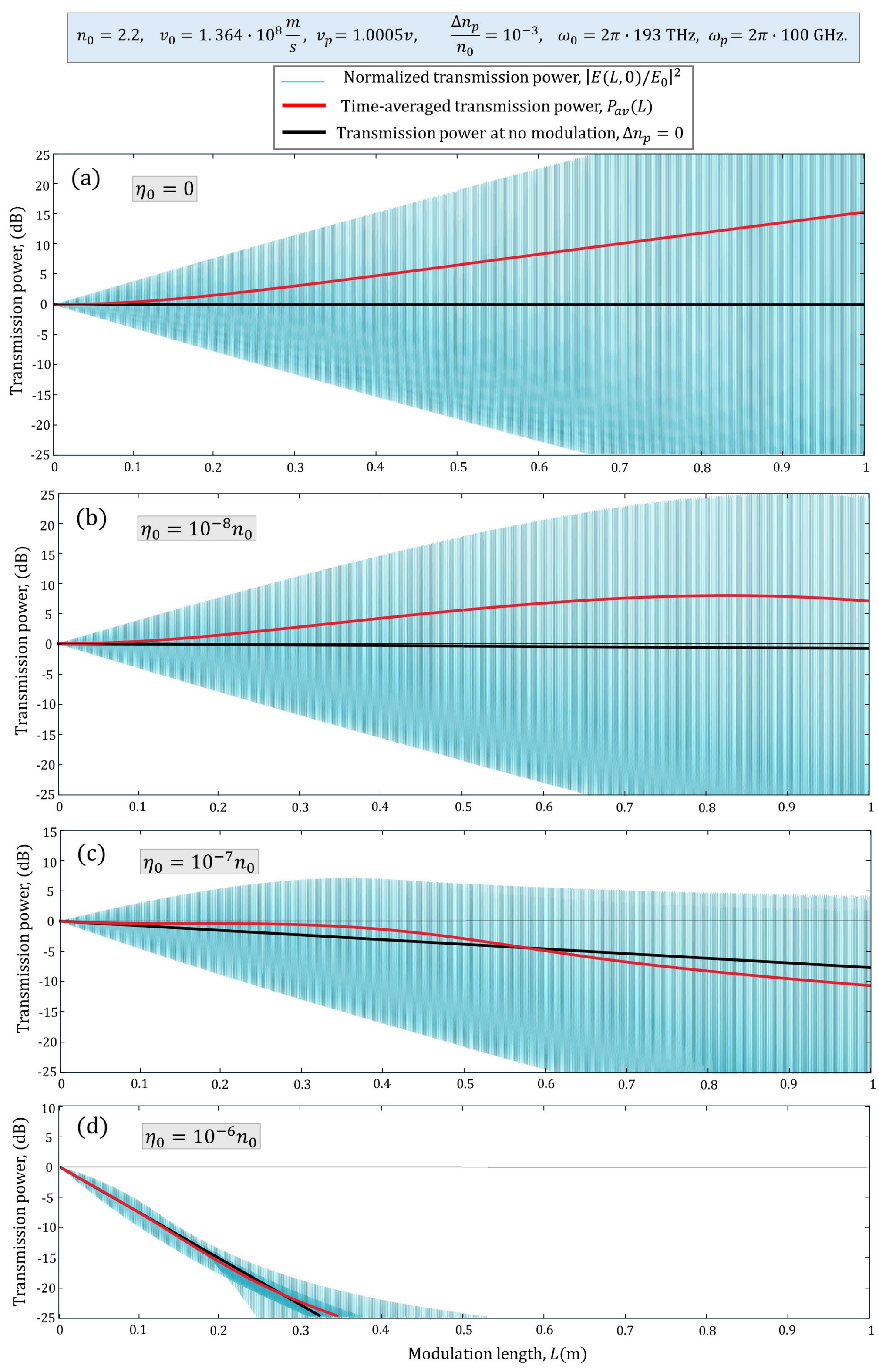

**Fig. 3.** The transmission power and time-averaged power of the output wave as a function of modulation length $L$ for different losses $\eta_0$ in the synchronous case $|\mu| = 0.2$, $v_p = 1.0005v$. (a) $\eta_0 = 0$. (b) $\eta_0 = 10^{-8}n_0$. (c) $\eta_0 = 10^{-7}n_0$. (d) $\eta_0 = 10^{-6}n_0$.

$$
\begin{aligned}
&E(x,t) = U_0(x,t)\exp\left(iS(x,t)\right),\\
&S(x,t) = \frac{2\omega}{\omega_p}\arctan\left(\frac{\tanh\left(\dfrac{\Delta n_p \omega_p x}{2n_0 v_0}\right) - \tan\left(\dfrac{\omega_p}{2}\left(t-\dfrac{x}{v_0}\right)\right)}{1-\tanh\left(\dfrac{\Delta n_p \omega_p x}{2n_0 v_0}\right)\tan\left(\dfrac{\omega_p}{2}\left(t-\dfrac{x}{v_0}\right)\right)}\right),\\
&U_0(x,t) = \frac{1}{\cosh\left(\dfrac{\Delta n_p \omega_p x}{n_0 v_0}\right) - \sin\left(\omega_p\left(t-\dfrac{x}{v_0}\right)\right)\sinh\left(\dfrac{\Delta n\ \omega_p x}{n_0 v_0}\right)}.
\end{aligned}
\tag{11}
$$

Here, the expression for $U_0(x,t)$ is derived under the commonly satisfied condition $\Delta n_p/n_0 \ll 1$. For a relatively small argument of cosh (...) in the expression for $U_0(x,t)$ in Eq. (11), $x\Delta n_p \omega_p/n_0 v_0 \ll 1$ (weak amplification), the expansion of $U_0(x,t)$ up to the second order in $x\Delta n_0 \omega_p/n_0 v_0$ coincides with that found in Ref. [16] where the case $|\mu| \ll 1$ was considered (see below). From Eqs. (10) and (11), the time-averaged wave power can be found analytically:

$$
P_{av}(L) = \cosh\left(\frac{\Delta n_p \omega_p L}{c}\right). \tag{12}
$$

This result coincides with that found for the exactly solvable problem of waveguides with constant impedance [8]. It is seen from Fig. 2 that the completely synchronous condition corresponds to the maximum wave power amplification. It follows from Eq. (12) and also seen in Fig. 2(f) that the average amplification power grows exponentially with modulation length $L$ if $L \gg c/\Delta n_p \omega_p$. For the optical waveguide and modulation parameters indicated in Fig. 3 we have $L \cong 0.5$ m. We find from Eq. (11) that for $x \gg c/\Delta n_p \omega_p$ the amplitude $U_0(x,t)$ becomes a fast function of the coordinate and time if $\sin\left(\omega_p\left(t - \frac{x}{v_p}\right)\right)$ is close to unity. In the latter case the eikonal approximation may fail since its condition of validity reads (see Appendix B):

$$
\frac{\omega_p}{\omega_0} \ll \exp\left(-\frac{2\Delta n_p \omega_p}{n_0 v_0}x\right). \tag{13}
$$

This condition is well satisfied for the parameters considered in our numerical modelling.

For a relatively small amplification length $L \ll c/\Delta n_p \omega_p$, the spectrum of the output light is localized near the input frequency $\omega_0$. Consequently, it is convenient to introduce the spectrum centered at $\omega_0$ by the expansion:

$$
E(L,t) = \sum_{m=-\infty}^{\infty} U_m^{(c)} \exp\left(-i(\omega_0 + m\omega_p)t\right), \tag{14}
$$

$$
U_m^{(c)} = \frac{\omega_p}{2\pi}\int_0^{2\pi/\omega_p} E(L,t)\exp\left(i(\omega_0 + m\omega_p)t\right)dt\,. \tag{15}
$$

For eikonal $S(x,t)$ and amplitude $U_0(x,t)$, which are slow functions of time, the integral for $U_n^{(c)}$ in Eq.

(15) can be calculated by the stationary phase method. Calculations detailed in Appendix C show that the stationary points of this integral exist only within the frequency band

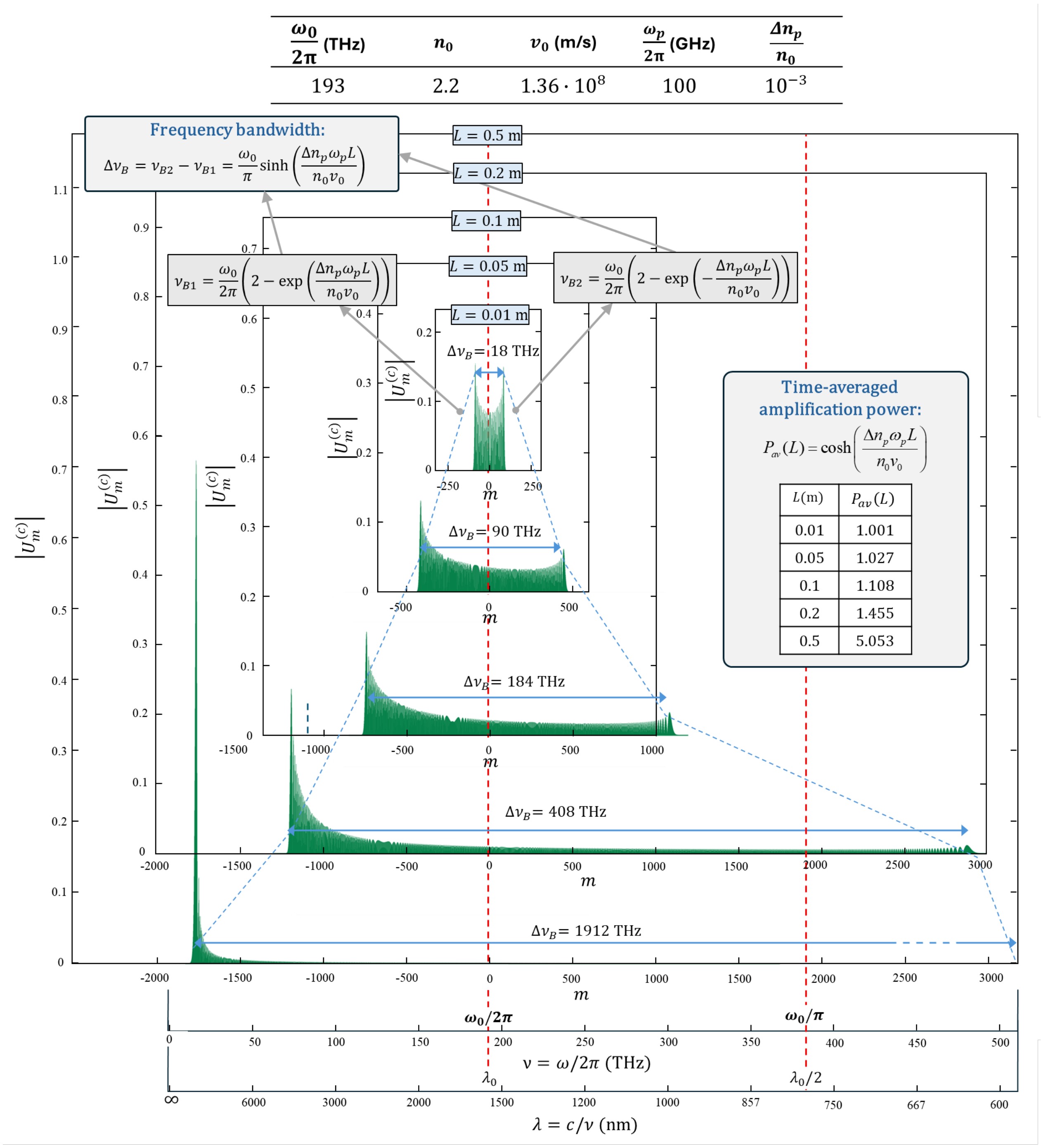

| $\frac{\omega_0}{2\pi}$ (THz) | $n_0$ | $v_0$ (m/s) | $\frac{\omega_p}{2\pi}$ (GHz) | $\frac{\Delta n_p}{n_0}$ |
|---|---|---|---|---|
| 193 | 2.2 | $1.36 \cdot 10^8$ | 100 | $10^{-3}$ |

| $L$(m) | $P_{av}(L)$ |
|---|---|
| 0.01 | 1.001 |
| 0.05 | 1.027 |
| 0.1 | 1.108 |
| 0.2 | 1.455 |
| 0.5 | 5.053 |

**Fig. 4.** Transmission spectra and time-average amplification power at the completely synchronous condition $v_p = v_0$ for different modulation lengths $L = 0.01, 0.05,\ 0.1, 0.2,$ and $0.5$ m. Parameters of the input light, waveguide, and modulation are indicated at the top of the figure.

$$\omega_{B1}(L) < \omega < \omega_{B2}(L), \quad \omega_{B1,2} = \omega_0 \left( 2 - \exp\left( \pm \frac{\Delta n_p \omega_p}{n_0 v_0} L \right) \right) \tag{16}$$

with the bandwidth

$$\Delta\omega_B = \omega_{B2}(L) - \omega_{B2}(L) = 2\omega_0 \sinh\left( \frac{\Delta n_p \omega_p}{n_0 v_0} L \right). \tag{17}$$

Close to the edges of this band, the second derivative $S_{xx}$ tends to zero and the stationary phase method fails. The absence of the real stationary points of the integral in Eq. (15) outside of this frequency band suggests that the position of the spectral bandwidth of the solution given by Eq. (11) is determined by Eqs. (16) and (17). This result is confirmed by numerical calculations of the spectrum for different modulation lengths $L = 0.01, 0.05,\ 0.1, 0.2,$ and 0.5 m presented in Fig. 4. For a relatively small modulation length $L = 0.01$ m, the average field amplification is negligible, $P_{av}(L) = 1.001$, and the spectral bandwidth is small compared to the input light frequency, $\Delta\nu_B = \Delta\omega_B/2\pi = 18$ THz $\ll \omega_0/2\pi = 193$ THz. At $L =$ 0.05 m and 0.1 m, the bandwidth $\Delta\nu_B$ becomes comparable to the input frequency, though the average amplification remains small. For larger modulation lengths $x_p$, the left-hand-side bandwidth edge $\omega_{B1}(L)$ becomes negative and exponentially grows with $L$. Alternatively, for large $L$ the right-hand-side edge $\omega_{B2}(L)$ tends to $2\omega_0$ and the output wave spectrum localizes in the vicinity of $2\omega_0$ as illustrated by the spectrum of the output wave at $L = 0.5$ m in Fig. 4.

The feasibility of amplification of light by a synchronous traveling wave can be better understood by comparing Eqs. (12) and (17). From these equations, we find a simple relation between the time-averaged normalized power and the spectral bandwidth of the outgoing wave:

$$P_{av}(L) = \sqrt{1 + \left( \frac{\Delta\omega_B(L)}{2\omega_0} \right)^2}\ . \tag{18}$$

Thus, significant time-averaged amplification is impossible if the spectrum bandwidth $\Delta\omega_B(L)$ is small compared to the input frequency $\omega_0$. Eq. (18) is derived under the condition of dispersionless propagation, which, for realistic optical waveguides, can be valid only within a relatively small bandwidth $\Delta\omega_B$. Since the transmission bandwidths of optical materials are also relatively small, the significant amplification of light in optical waveguides parametrically modulated by a traveling wave is currently unfeasible.

**E. Wave propagation and wave spectrum in a completely asynchronous case $|\mu| \ll 1$**

In contrast to the amplification, the traveling wave modulation with the phase velocity $v_p$ approaching the phase velocity of light $v_0$ is important for the enhancing the performance of broadband optical modulators and frequency comb generators [19-28]. The frequency comb bandwidth, which is close to maximum possible for a given modulation length $L$ and amplitude $\Delta n_p$, can be achieved without the accurate proximity to the completely synchronous condition $v_p = v_0$ (i.e., satisfying the condition $|\mu| \gg 1$ and even $|\mu| > 1$) considered in the previous section. Here we demonstrate these results considering the asynchronous case

$$|\mu| = \frac{\Delta n_p v_p}{n_0 |v_p - v_0|} \ll 1. \tag{19}$$

Keeping the zero and the first order in $|\mu|$ and $\eta_0$ terms in the expression for the wave amplitude $U_0(x,t)$ and phase $S(x,t)$ given by Eqs. (6)-(9), we find:

$$E(x,t) = U_0(x,t)\exp\left(iS(x,t)\right),$$

$$S(x,t) = \omega_0\left(\frac{x}{v_0} - t\right) + i\frac{\omega_0\eta_0}{c}x + \Omega_p(x)\cos\left(\omega_p\left(t - \frac{\left(v_0 + v_p\right)}{2v_0 v_p}x\right)\right), \tag{20}$$

$$U_0(x,t) = 1 - \Omega_p(x)\frac{\omega_p}{2\omega_0}\left(\frac{v_0}{v_p} - 3\right)\sin\left(\omega_p\left(t - \frac{\left(v_0 + v_p\right)}{2v_0 v_p}x\right)\right).$$

In this equation, we introduce the *modulation index* $\Omega_p(x)$ which characterizes the effect of modulation on the propagation of a wave along the modulation length $L$:

$$\Omega_p(L) = \frac{2\Delta n_p \omega_0 v_p}{n_0\omega_p\left(v_0 - v_p\right)}\sin\left(\frac{\left(v_0 - v_p\right)}{2v_0 v_p}\omega_p L\right). \tag{21}$$

The structure of the eikonal $S(x,t)$ in Eq. (20) resembles the expressions known in the theory of optical modulators [20, 24]. In particular, it follows from this equation that, in this case, the effect of material losses is described by the factor exp $(-\omega_0\eta_0 x/c)$, which is the same as for the stationary (unmodulated) wave propagation. For the case of instantaneous modulation, $v_p = \infty$, Eq. (21) coincides with that for the absolute value of modulation index found in Ref. [58]. The expression for the normalized amplitude $U_0(x,t)$ in Eq. (20) shows that its deviations from unity is commonly small due to the factor $\omega_p/\omega_0 \ll 1$. It also follows from this expression that the first order in modulation amplitude $\Delta n_p$ term in the time-averaged power $P_{av}(x)$ defined by Eq. (10) vanishes. Taking into account the second order in $\Delta n_p$ (more precisely – in $\mu$) term, we find:

$$\begin{aligned} P_{av}(L) &\cong 1 - \frac{2\omega_0\eta_0}{c}x + \frac{\left(v_0 - 2v_p\right)\left(v_0 - 3v_p\right)}{v_p^2}\left(\Omega_p(L)\frac{\omega_p}{\omega_0}\right)^2 \\ &= 1 - \frac{2\omega_0\eta_0}{c}L + \frac{\Delta n_0^2}{4n_0^2}\frac{\left(v_0 - 2v_p\right)\left(v_0 - 3v_p\right)}{\left(v_0 - v_p\right)^2}\sin^2\left(\frac{v_0 - v_p}{2v_0 v_p}\omega_p L\right). \end{aligned} \tag{22}$$

This equation shows that, close to the completely synchronous condition $v_0 = v_p$ or, more precisely, for $|v_0 - v_p| \ll v_0$ and modulation lengths satisfying the inequality

$$L \ll L_s = \frac{2v_0 v_p}{\omega_p\left|v_0 - v_p\right|}, \tag{23}$$

the modulation leads to a small wave amplification equal to

$$P_{av}(L)\big|_{L<<L_s} \cong 1 - \frac{2\omega_0\eta_0}{c}L + \frac{\Delta n_0^2\omega_p^2}{2c^2}L^2. \tag{24}$$

Unexpectedly, for zero losses, $\eta_0 = 0$, this result coincides with that given by Eq. (12) for $\Delta n_0 \omega_p L/c \ll 1$. This result is also similar to that found in Ref. [16] under the same assumption $|\mu| \ll 1$. From Eq. (24), the modulation length leading to the noticeable amplification of light can be estimated as $L_a = c/(\Delta n_0 \omega_p)$. For the parameters considered here, $\omega_p \sim 2\pi \cdot 100$ GHz, $n_0 = 2.2$, $\Delta n_0/n_0 \sim 10^{-3}$, we have $L_a \sim 20$ cm. Eq. (22) shows that $P_{av}(L)$ vanishes if the relation between phase velocities is close to $v_0 = 2v_p$ and $v_0 = 3v_p$. Alternatively, modulation leads to the wave attenuation if $2v_p < v_0 < 3v_p$. It follows from Eq. (22) that amplification is always small outside the vicinity where $|v_0 - v_p|/v_p \ll 1$. This result is also evident from the solutions given by Eq. (6)-(9) for $|\mu| \ll 1$ and $\Delta n_p/n_0 \ll 1$. Eq. (22) will be used below to discuss the relation between the possible amplification and the required transmission bandwidth.

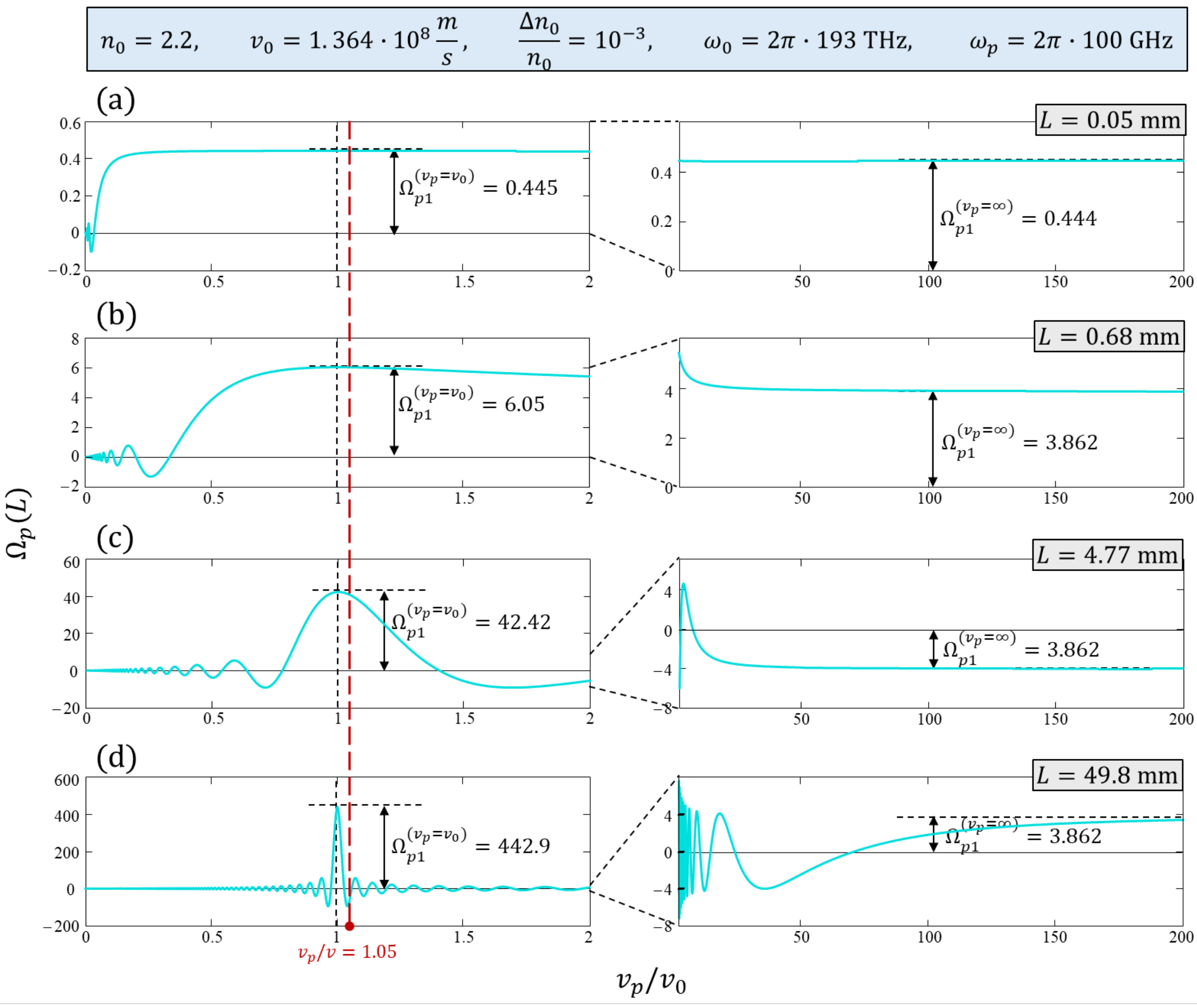

**Fig. 5.** Modulation index $\Omega_p$ as a function of the ratio $v_p/v_0$ at different modulation lengths, (a) $L = 0.05$ mm, (b) $L = 0.68$ mm, (c) $L = 4.77$ mm, and (d) $L = 49.8$ mm, for the lithium niobate waveguide. The light and modulation parameters are shown at the top of the figure.

The characteristic dependencies of $\Omega_p(L)$ as a function of the ratio $v_p/v$ at different modulation lengths, $L = 0.05$ mm, 0.68 mm, 4.77 mm, and 49.8 mm, are shown in Fig. 5. In this figure, we again assume that the waveguide refractive index is that of lithium niobate, $n_0 = 2.2$, the relative amplitude of refractive index modulation is $\Delta n_p/n_0 = 10^{-3}$, and the light and modulation frequencies are $\omega_0 = 2\pi \cdot 193$ THz and $\omega_p = 2\pi \cdot 100$ GHz. Due to the periodic dependence of $\Omega_p(L)$, the values of $L$ in Fig. 5 correspond to the maxima of $\Omega_p(L)$ nearest to the lengths $L = 0.05$, 0.5, 5, and 50 mm at $v_p = \infty$. From Eq. (21), these maxima are situated at the periodic sequence of modulation lengths $L = \frac{\pi v_0}{\omega_p}(2N+1)$, $N = 0, 1, 2 ....$ It is seen from the plots of Fig. 5 that, while the modulation index for the completely synchronous modulation ($v = v_p$) and instantaneous modulation ($v_p = \infty$) are close to each other for the small modulation length $L \lesssim 0.1$ mm, the modulation index becomes much greater for larger modulation lengths at $v_p$ approaching $v$. Due to the condition of Eq. (19), the synchronous case and, in particular, the exact equality $v = v_p$ is excluded in the considered approximation. However, as follows from Eq. (19), the plots in Fig. 5 are accurate everywhere except for a relatively small vicinity of the completely synchronous coordinate $v_p/v = 1$ where $|v_p/v_0 - 1| \sim \Delta n_p/n_0 = 10^{-3}$. Therefore, these plots are reasonably accurate for all $v_p/v_0$. Assuming that the modulation length $L$ is sufficiently small as defined by Eq. (23), or that $v_p/v_0 \to 1$, we simplify Eq. (21) for $\Omega_p(L)$ to

$$\Omega_p(L)\Big|_{L<<L_s} = \frac{\Delta n_p \omega_0}{c} L . \tag{25}$$

From this equation, the maximum of modulation index at $v_0 \cong v_p$ grows linearly with the modulation length $L$. Indeed, Fig. 5 shows that the modulation index can be dramatically increased for large $L$ if the phase velocity of the traveling wave is sufficiently close to the phase velocity of light.

While, for realistic waveguides, the substantial amplification of light is unfeasible, the modulation index $\Omega_p(L)$ defined by Eq. (21) can significantly exceed unity and lead to the creation of a relatively broadband comb spectrum near the synchronous condition $v_p = v_0$ (see Fig. 5(c) and (d)). To determine the spectrum of the output wave for $|\mu| \ll 1$, we rewrite Eq. (20) as

$$E(x,t) = \exp\left[ i\omega_0 \left( \frac{x}{v_0} - t \right) - \frac{\omega_0 \eta_0}{c} x + i\Omega_p(x) \cos\left( \omega_p \left( t - \frac{v_0 + v_p}{2 v_0 v_p} x \right) - iG_p \right) \right]. \tag{26}$$

Here we introduce a small parameter

$$G_p = \frac{\omega_p}{2\omega_0} \left( \frac{v_0}{v_p} - 3 \right), \quad |G_p| \ll 1. \tag{27}$$

We note that $|G_p| \ll 1$ due to Eq. (5). Applying the Jacobi-Unger expansion to Eq. (26) we find:

$$\begin{aligned} E(L,t) &= \sum_{m=-\infty}^{\infty} U_m^{(c)} \exp\left[ -i\left( \omega_0 + m\omega_p \right) t \right], \\ U_m^{(c)} &= J_m\left( \Omega_p(L) \right) \exp\left[ -\frac{\omega_0 \eta_0}{v_0} L - \frac{i\pi m}{2} + i\omega_0 \frac{L}{v_0} + i\omega_p \frac{v_0 + v_p}{2 v_0 v_p} Lm - G_p m \right]. \end{aligned} \tag{28}$$

To estimate the maximum possible amplitude of conversion $\omega_0 \to \omega_0 + m\omega_p$, we note that, while the maximum argument of the Bessel function in Eq. (28) can be large, $z = \Omega_p(L) \gg 1$, the maximum of $|J_m(z)|$ is always smaller than $2^{-1/2}$ [59]. For $|m| \gg 1$, the $|J_m(z)|$ maximum is defined by its asymptotics equal to $0.674|m|^{-1/3}$ [59], i.e., vanishes very slowly. This maximum is achieved at $z \cong |m|$, while $|J_m(z)|$ rapidly vanishes for $|z| > |m|$. Thus, the frequency comb bandwidth of $E(x_p, t)$ is determined from Eq. (28) as

$$\Delta\omega_B(L) = 2\omega_p \left|\Omega_p(L)\right| , \tag{29}$$

and the amplification of a comb line is determined from the Eqs. (27) and (28) by the factor

$$F_m = \exp\left(-mG_p\right) = \exp\left[-m\frac{\omega_p}{2\omega_0}\left(\frac{v_0}{v_p} - 3\right)\right]. \tag{30}$$

Since in the approximation considered $\left|G_p\right| \ll 1$, the factor $F_m$ can be large only for sufficiently large negative comb line numbers $m$. From Eqs. (29) and (30), we find the maximum possible amplification takes place for the maximum absolute value of negative $m = -|\Omega_p(L)|$ within this bandwidth. Then, we find from Eq. (21) for $\Omega_p(L)$ that the amplification effect defined by $F_m$ is always small away from the synchronous condition, confirming the general result directly following from Eq. (22).

As noted above, proximity to the completely synchronous condition $v_p = v_0$ can significantly increase the modulation index (up to the values of ~ 40 and ~ 400 for the lithium niobate waveguide with $L \cong 5$ mm and $L \cong 50$ mm, respectively, see Fig. 5) and as follows from Eq. (29) can increase the frequency comb bandwidth $\Delta\omega_p$ proportionally. A dramatic enhancement of the spectral bandwidth generated by a traveling wave having the phase velocity close to $v_0$ though still for $|\mu| \ll 1$, as compared to the bandwidth generated by the instantaneous modulation with $v_p = \infty$ and reverse modulation with the reverse sign of $v_p$, is evidenced from Fig. 6. The reason for the enhancement is a much greater value of the modulation index of a traveling wave having $v_p \cong v_0$. The parameters of light, waveguide, and modulation, which are similar to the parameters considered in our previous examples, are indicated at the top of this figure. For the waveguide with these parameters, the value of $\Omega_p(L)$ achieves 81 at $L = 14.3$ mm (Fig. 6(a)) at $v_p = 1.05v$, though it remains much smaller for the instantaneous and reverse modulations when $v_p = \infty$ and $v_p = -1.05v_0$, respectively. Figure 7(b) shows the dependences of the reduced eikonal, $S(L, 0) - \omega_0 L/v_0$, on the modulation length for the same phase velocity relations as well as for the completely synchronous case $v_p = v_0$. It is seen that the oscillation amplitude of the eikonal as a function of modulation length is proportional to the local modulation index, as directly follows from Eq. (21). Figures 7(c), (d), and (e) compare the frequency comb spectra of solutions for (c) $v_p = 1.05v_0$, (d) $v_p = \infty$, and (e) $v_p = -1.05v$ for different modulation lengths $L$ indicated in the plot (b) of Fig. 6. It is seen that, in accordance with Eq. (29), the generated frequency comb bandwidth is proportional to the value of the corresponding modulation index shown in Fig. 6(a). We note that the dramatic reduction of the modulation effect of the reverse vs. the direct modulation manifests the strong nonreciprocity of the considered device.

In contrast to the optical frequency comb bandwidth, the total amplification of light induced by modulation remains small for $\Delta\omega_B(L) \ll \omega_0$. Indeed, comparing Eq. (29) and Eq. (22), we find:

$$P_{av}(L) \cong 1 - \frac{2\omega_0\eta_0}{c}L + \frac{\left(v_0 - 2v_p\right)\left(v_0 - 3v_p\right)}{v_p^2}\left(\frac{\Delta\omega_B(L)}{4\omega_0}\right)^2 . \tag{31}$$

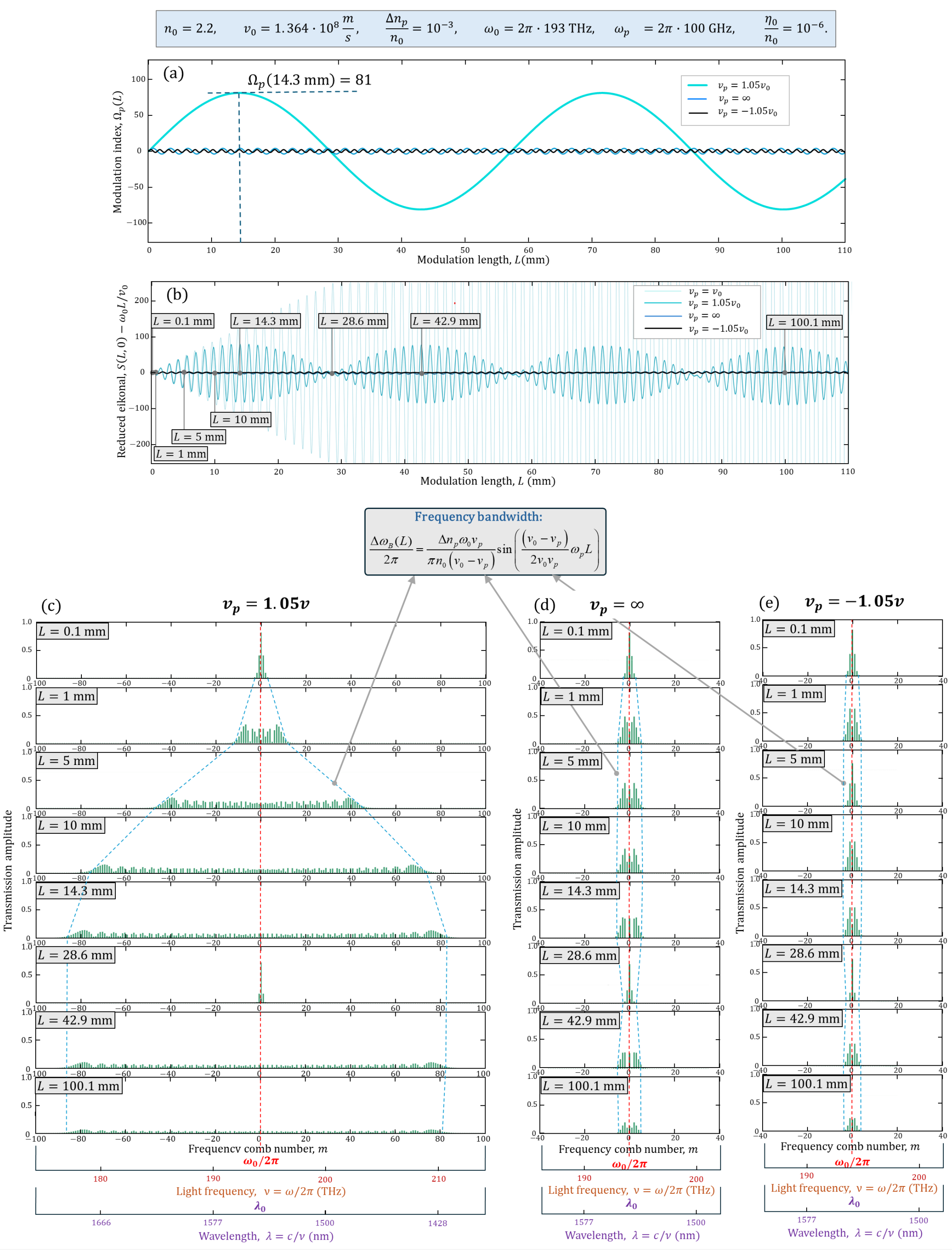

**Fig. 6.** (a) Dependence of the modulation index $\Omega_p(L)$ on the modulation length $L$ for close phase velocities, $v_p = 1.05v$, in the asynchronous case $|\mu| = 0.02$ (light blue curve), for the instantaneous modulation, $v_p = \infty$ (blue curve), and for the reverse modulation $v_p = -1.05v_0$ (black curve). (b) Dependences of the reduced eikonal, $S(L,0) - \omega_0 L/v_0$, on the modulation length $L$ for the completely synchronous modulation, $v_p = v_0$ (dimmed light blue curve), $v_p = 1.05v_0$ (light blue curve), $v_p = \infty$ (blue curve), and $v_p = -1.05v_0$ (black curve). (c), (d), and (e) The transmission amplitudes for (c) $v_p = 1.05v_0$, (d) $v_p = \infty$, and (d) $v_p = -1.05v_0$ for different modulation lengths $L$ indicated in the plots of this figure.

In particular, close to the completely synchronous condition $v_p = v_0$

$$P_{av}(L)\big|_{v_p \to v_0} = 1 - \frac{2\omega\eta_0}{c} L + \frac{1}{8}\left(\frac{\Delta\omega_B(L)}{\omega_0}\right)^2 . \tag{32}$$

Remarkably, this equation coincides with Eq. (18) for a relatively small bandwidth $\Delta\omega_B \ll \omega_0$. Assuming that velocities $v_0$ and $v_p$ are of the same order, we find from Eq. (31) that, similar to the completely synchronous case described by Eq. (18), significant amplification of light is impossible in realistic waveguides, which always have $\Delta\omega_B \ll \omega_0$. However, the situation for small traveling wave velocities $v_p \ll v_0$ cannot be clarified from Eq. (31) due to the restriction of the eikonal approximation, $v_p \gg v_0\, \omega_p/\omega_0$ following from Eq. (5). The latter restriction will be removed in the following Section.

## III. THE PERTURBATION THEORY APPROACH

The eikonal approximation used above does not allow us to consider sufficiently small values of $v_p$ since, according to Eq. (5), the condition of slowness of modulation in space restricts these values to $v_p \gg \omega_p v_0/\omega_0$. However, we will show below in Section IV that the case of comparable $\omega_p/v_p \sim \omega_0/v_0$ is important to arrive at the strong amplification of light in an optical resonator. We find from Eq. (21) that for $\omega_p/v_p \sim \omega_0/v_0$ and $\omega_p \ll \omega_0$ the modulation index $|\Omega_p(L)| \sim \Delta n_p/n_0 \ll 1$. Under the latter condition, the restriction $\omega_p/v_p \ll \omega_0/v_0$ can be withdrawn and solution of the wave equation, Eq. (1), can be found by the regular perturbation theory.

Having in mind modulation by acoustic traveling waves, which for a relatively large IDT tilt angle $\theta$ (see Fig. 1(c)) may strongly attenuate in space [60, 61], we consider now the refractive index in Eq. (1) in the form (Fig. 7):

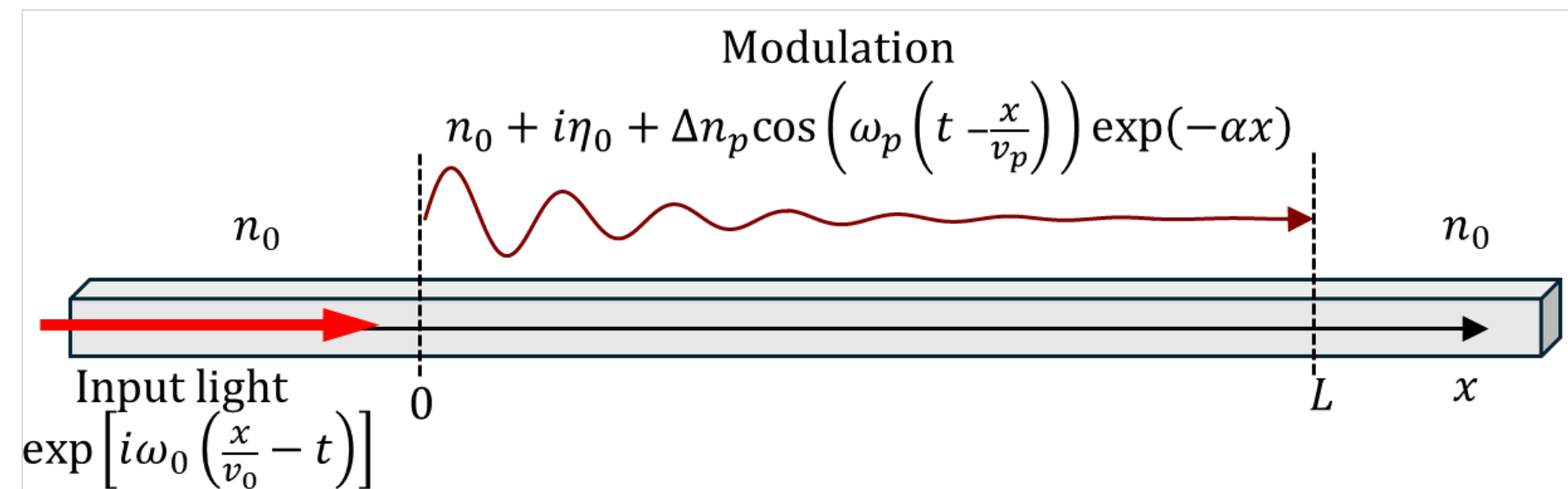

**Fig. 7.** An optical waveguide with the refractive index modulated by a traveling wave fully attenuating within the interval $(0, L)$.

$$\begin{aligned} n(x,t) &= n_0 + \Delta n(x,t), \\ \Delta n(x,t) &= i\eta_0 + \Delta n_p \cos\left(\omega_p\left(t - \frac{x}{v_p}\right)\right)\exp\left(-\alpha_p x\right) \end{aligned} \tag{33}$$

Here $\alpha_p$ defines the attenuation of the modulating wave. In the first order over $\Delta n_p/n_0$ and $\eta_0/n_0$, the solution of Eq. (1) with the boundary condition

$$E^{(0)}(x,t) = \exp\left( i\omega_0 \left( \frac{x}{v_0} - t \right) \right) \tag{34}$$

is found as

$$E(x,t) = E^{(0)}(x,t)\left(1 + \Delta U_p(x,t)\right), \tag{35}$$

where $\left|\Delta U_p(x,t)\right| \ll 1$. Under the latter condition, it is convenient to present this solution in the form similar to that in Eq. (26) using $1 + \Delta U_p(x,t) \cong \exp\left(\Delta U_p(x,t)\right)$. Then the calculations detailed in Appendix D yield:

$$E(x,t) = E^{(0)}(x,t)\exp\left( -\eta_0 \frac{\omega_0 x}{c} + i\tilde{\Omega}_p(x)\cos\left( \omega_p \left( t - \frac{v_0 + v_p}{2v_0 v_p} x \right) - i\tilde{G}_p(x) \right) \right), \tag{36}$$

$$\tilde{\Omega}_p(x) = -2i\sqrt{\Delta U^+ \Delta U^- W^+(x) W^-(x)}, \quad \tilde{G}_p(x) = \frac{1}{2}\ln\left( \frac{\Delta U^- W^-(x)}{\Delta U^+ W^+(x)} \right), \tag{37}$$

where

$$W^{\pm}(x) = \sin\left( \pm \frac{v_0 - v_p}{2v_0 v_p}\omega_p x + \frac{i\alpha_p}{2} x \right)\exp\left( -\frac{\alpha_p}{2} x \right), \tag{38}$$

$$\Delta U^{\pm} = \frac{\Delta n_p v_p^2 \left(\omega_0 \pm \omega_p\right)^2}{n_0 \left[ \pm\omega_p \left(v_0 - v_p\right) + i\alpha_p v_0 v_p \right]\left[ 2\omega_0 v_p \pm \omega_p \left(v_0 + v_p\right) + i\alpha_p v_0 v_p \right]}. \tag{39}$$

This solution is valid only if $\left|\Delta U_p(x,t)\right| \ll 1$, or

$$\left|\Delta U^{\pm}\right| \ll 1. \tag{40}$$

Obviously, in contrast to the eikonal approximation, the solution described by Eqs. (36)-(39) includes transitions with acquisition or loss of a single frequency $\omega_p$ only. This solution includes a backward propagating wave if the wavenumber $k_p$ of the traveling wave is larger than the wavenumber $k_0$ of the input wave, while the perturbed wave is forward propagating in the opposite case, $k_p < k_0$ (see Appendix D). Here, we restrict our consideration to the forward propagation case, $k_p < k_0$, and chose the solution so that it vanishes at the starting point of modulation, $x = 0$, $\Delta U_p(0,t) = 0$. The important case $k_p > k_0$ including Brillouin backscattering will be considered elsewhere [62]. At zero attenuation, $\alpha_p = 0$, under the conditions of Eq. (40) and Eq. (5), the determined solution coincides with that given by Eq. (26). In particular, $\tilde{\Omega}_p(L)$ coincides with $\Omega_p(L)$ and $\tilde{G}_p(L)$ coincides with $G_p$.

## IV. TRANSFORMATION AND AMPLIFICATION OF LIGHT BY AN OPTICAL RESONATOR

We consider now a closed optical waveguide with length $2L$ forming a racetrack resonator which is coupled to an input-output waveguide as illustrated in Fig. 8. We assume that the modulation is described by Eq. (1) with the refractive index defined by Eq. (2) (Section IV.A) and by Eq. (33) (Section IV.B) and takes place along the length $L$ of the resonator waveguide. The monochromatic input light in the input-output waveguide near position $x = x_0$ in front of the coupling region is set to $E_{in}(x,t) = \exp\big(i\omega_0(x/v_0 - t)\big)$. As noted above, here we consider only the case of *forward propagating optical waves*, i.e., assume that $k_p < k_0$ and modulation does not introduce backward propagation of light. In addition, we assume that the input light power is small enough so that the nonlinear effects caused by the resonance propagation of light in the resonator are negligible. Then, using the transfer matrix approach (see, e.g., [63]), we find the output light field $E_{out}(t)$ from the equation

$$\begin{pmatrix} E_{out}(t) \\ E(x_0,t) \end{pmatrix} = S \begin{pmatrix} E_{in}(t) \\ E(2L + x_0, t) \end{pmatrix}, \quad S = \begin{pmatrix} \tau & \kappa \\ -\kappa & \tau \end{pmatrix}, \tag{46}$$

where matrix $S$ is the unitary S-matrix, so that $\tau^2 + \kappa^2 = 1$. In this equation, the coordinates $x = 2L + x_0$ and $x = x_0$ define the beginning and the end of the coupling region and the S-matrix parameters $\kappa$ and $\tau$ determine the coupling between the input-output waveguide and resonator (Fig. 8).

### A. The eikonal approximation

In our calculations, we follow the approach of Ref. [58] where the determination of the output wave $E_{out}(t)$ was reduced to the solution of a functional equation. For modulation without attenuation (Fig. 8(a)), using Eq. (20) we find:

$$E(x_0,t) = \exp(-i\omega t)\Phi(t), \tag{47}$$

$$E(2L + x_0, t) = \exp(-i\omega t) A(t) \Phi(t - T), \quad T = \frac{2L}{v_0}, \tag{48}$$

$$A(t) = \exp\left[ i\omega_0 T - \frac{\eta_0}{n_0}\omega_0 T + i\Omega_p(L)\cos\left( \omega_p \left( t - \frac{(v_0 + v_p)}{2v_0 v_p} L - iG_p \right) \right) \right]. \tag{49}$$

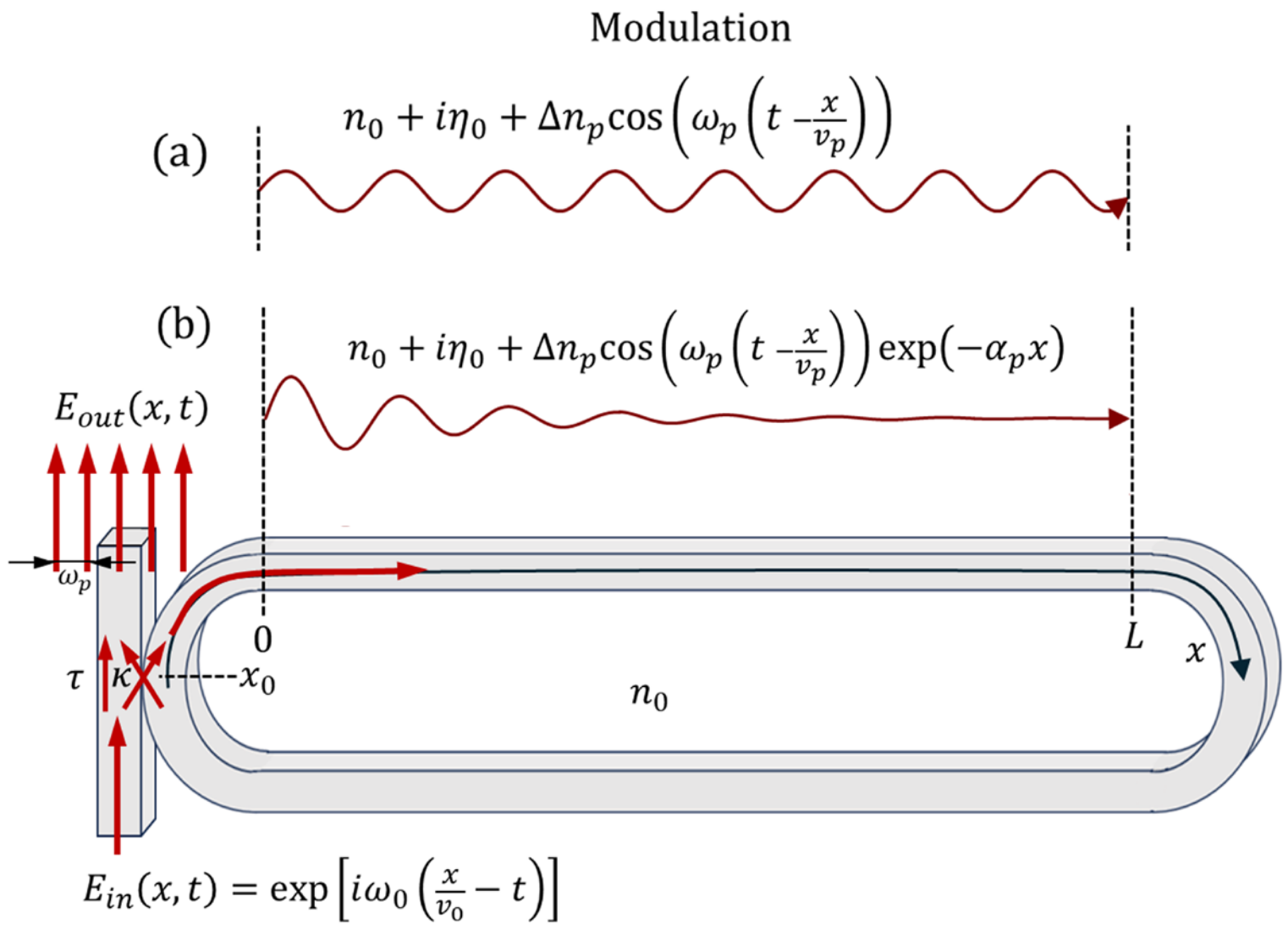

**Fig. 8.** An optical racetrack resonator with the refractive index modulated by a traveling wave along the interval $(0, L)$. (a) Uniform modulation. (b) Attenuating modulation.

Here $T$ is the roundtrip circulation time. Eqs. (46)-(49) lead to the functional equation for the arbitrary function $\Phi(t)$:

$$\Phi(t)=\tau A(t)\Phi\left(t-T\right)-\kappa \tag{50}$$

which can be solved exactly [58]. As a result, in full analogy with calculations of Ref. [58] (see Appendices B and C of [58]), the comb spectral amplitudes $U_m^{(c)}$ of $E_{out}(t)$ are found from the expansion:

$$\begin{gathered}
E_{out}(t)=\sum_{m=-\infty}^{\infty}U_m^{(c)}\exp\left[-i\left(\omega_0+m\omega_p\right)t\right],\\
U_m^{(c)}=\tau\delta_{0m}-\kappa^2\exp\left[im\left(-\frac{\pi}{2}+\frac{\omega_p T}{2}+\frac{\omega_p\left(v_0+v_p\right)}{2v_0v_p}L+iG_p\right)\right]\times\\
\sum_{n=0}^{\infty}\tau^n J_m\left(\sigma_{n+1}\Omega_p(L)\right)\exp\left[(n+1)\left(\frac{im}{2}\omega_p T+i\omega_0 T-\frac{\eta_0}{n_0}\omega_0 T\right)\right],\\
\sigma_n=\frac{\sin\left(\frac{n}{2}\omega_p T\right)}{\sin\left(\frac{1}{2}\omega_p T\right)},\qquad G_p=\frac{\omega_p}{2\omega_0}\left(\frac{v_0}{v_p}-3\right),
\end{gathered} \tag{51}$$

where $\delta_{nm}$ is the Kronecker delta and the amplification parameter $G_p$ is the same as in Eq. (27). The total time-averaged output power is calculated from this equation as

$$P_{av}^{(out)}=\sum_{m=-\infty}^{\infty}\left|U_m^{(c)}\right|^2. \tag{52}$$

In our further calculations, we assume that the coupling between the input-output and resonator waveguides $\kappa$ is small so that $\tau\cong 1-\kappa^2/2$ and, in Eq. (51), $\tau^n\cong\exp\left(-n\kappa^2/2\right)$. Under this assumption, the microresonator Q-factor found from Eq. (51) is

$$Q=\frac{1}{2}\left(\frac{\eta_0}{n_0}+\frac{\kappa^2}{2\omega_0 T}\right)^{-1} \tag{53}$$

and its intrinsic Q-factor is $Q_{int}=n_0/(2\eta_0)$.

Similar to Eq. (28), which describes the nonresonant propagation, the difference of the expressions for the spectral amplitudes $U_m^{(c)}$ determined here for the traveling wave modulation compared to those previously found for the instantaneous modulation [58], consists in a different expression for the modulation index $\Omega_p(L)$ defined now by Eq. (21), the additional phase factor $\exp\left(-im\omega_p(v_p+v_0)L/(2v_0v_p)\right)$, and amplification factor $F_m$ defined by Eq. (30). Now, in contrast to the nonresonant propagation, the factor $F_m$, which can be large for a sufficiently small traveling wave phase velocity $v_p$ and large negative comb line numbers $m$. Therefore, this factor can significantly increase the total output power as well as the power of individual frequency comb lines.

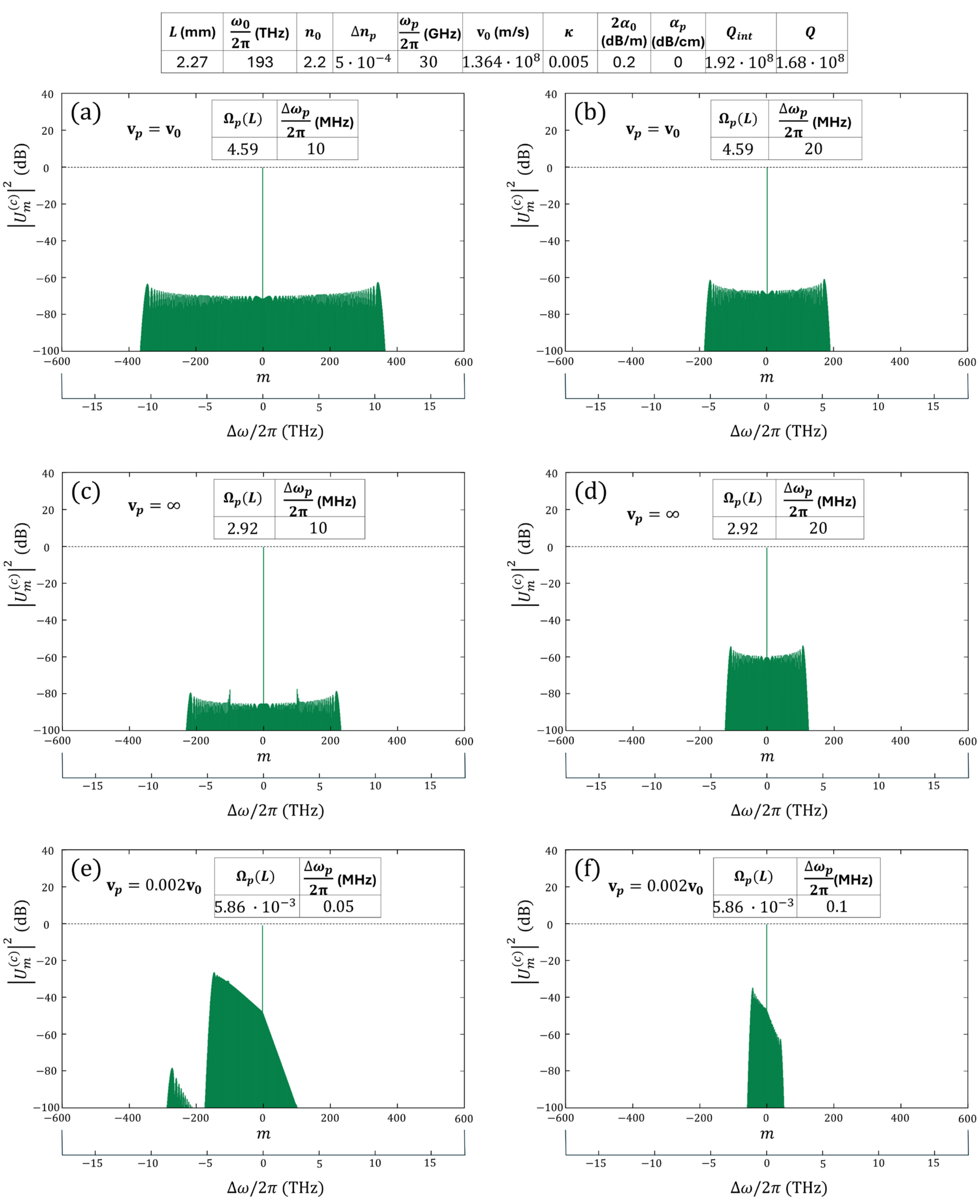

| $L$ (mm) | $\frac{\omega_0}{2\pi}$ (THz) | $n_0$ | $\Delta n_p$ | $\frac{\omega_p}{2\pi}$ (GHz) | $v_0$ (m/s) | $\kappa$ | $2\alpha_0$ (dB/m) | $\alpha_p$ (dB/cm) | $Q_{int}$ | $Q$ |
|---|---|---|---|---|---|---|---|---|---|---|
| 2.27 | 193 | 2.2 | $5 \cdot 10^{-4}$ | 30 | $1.364 \cdot 10^8$ | 0.005 | 0.2 | 0 | $1.92 \cdot 10^8$ | $1.68 \cdot 10^8$ |

**Fig. 9.** The resonant transmission power spectra for a racetrack resonator. The system parameters are indicated at the top of the figure. Plots (a)-(f) correspond to different relations between the velocity of light $v_0$ and traveling wave velocity $v_p$: (a), (b) $v_p = v_0$; (c), (d) $v_p = \infty$; and (e), (f) $v_p = 0.002v_0$. The right-hand side plots vs. the left-hand side plots demonstrate the shrinking of the transmission with the deviation $\Delta\omega_p$ from the exact modulation resonance condition.

The maximum amplitude and bandwidth of the output wave spectrum defined by Eq. (51) is achieved at the exact optical and modulation resonances, $\omega_0 = \omega_{0,N_0}^{(res)}$ and $\omega_p = \omega_{p,N_p}^{(res)}$ determined, respectively, by equations:

$$\omega_{0,N_0}^{(res)} T = 2\pi N_0, \quad N_0 \gg 1, \text{ integer}, \tag{54}$$

and

$$\omega_{p,N_p}^{(res)} T = 2\pi N_p, \quad N_p = 1, 2, \ldots . \tag{55}$$

It follows from the expression for $U_m^{(c)}$ in Eq. (51) that the deviation of the modulation frequency from this resonance condition,

$$\Delta\omega_p = \omega_{p,N_p}^{(res)} - \frac{2\pi N_p}{T}, \tag{56}$$

will reduce the magnitude of $U_m^{(c)}$. Indeed, the reduction grows with the frequency comb number $m$ due to the term $i(n+1)m\omega_p T/2$ in the exponent of the sum over $n$. This result, illustrated in Fig. 5(b) of Ref. [64] for the instantaneous modulation ($v_p = \infty$), suggests that choosing an appropriate offset $\Delta\omega_p$ we can appropriately shrink the transmission bandwidth $\Delta\omega_B$ of the resonator.

The effect of the modulation frequency offset $\Delta\omega_p$ on the transmission bandwidth and light amplification is illustrated in Fig. 9. In this figure, we again assume that the waveguide refractive index is that of lithium niobate, $n_0 = 2.2$, and set the light and modulation frequencies equal to $\omega_0 = 2\pi \cdot 193$ THz and $\omega_p = 2\pi \cdot 30$ GHz, the amplitude of refractive index modulation equal to $\Delta n_p = 5 \cdot 10^{-4}$, and waveguide loss equal to $2\alpha_0 = 0.2$ dB/m (the smallest value demonstrated to date []). It is seen from Fig. 9(a) that, for the synchronous modulation when $v_p = v_0$, significant offset values $\Delta\omega_p \sim 2\pi \cdot 10$ MHz are required to confine the output light within the bandwidth ~ 10 THz. Comparison of Figs. 9(a) and 10(b) shows that increasing the offset $\Delta\omega_p$ allows one to decrease the bandwidth proportionally to $\Delta\omega_p^{-1}$. However, no significant growth of the optical frequency amplitudes with shrinking the bandwidth are observed. In contrast, in agreement with Ref. [64], Figs. 9(b) and (c) show that, for the instantaneous modulation, $v_p = \infty$, these amplitudes significantly grow with $\Delta\omega_p$ and, consequently, with the reduction of the optical comb bandwidth. Figs. 9(e) and 10(f) consider the case of a relatively small phase velocity of the modulation wave, $v_p = 0.002 v_0$. In this case, the effect of the amplification factor $F_m = \exp(-mG_p)$, where $G_p$ grows proportionally to $v_0/v_p$ (see Eqs. (51) and (30)), becomes significant and shows up in the negative tilting of the comb spectrum. However, the validity of the eikonal approximation used, $\frac{v_p}{v_0} \gg \frac{\omega_p}{\omega_0} = 1.5 \cdot 10^{-4}$ (Eq. (5)), does not allow us to consider smaller modulation wave velocities. In Section IV.B, we will overcome this limitation using the perturbation theory approach developed in Section III.

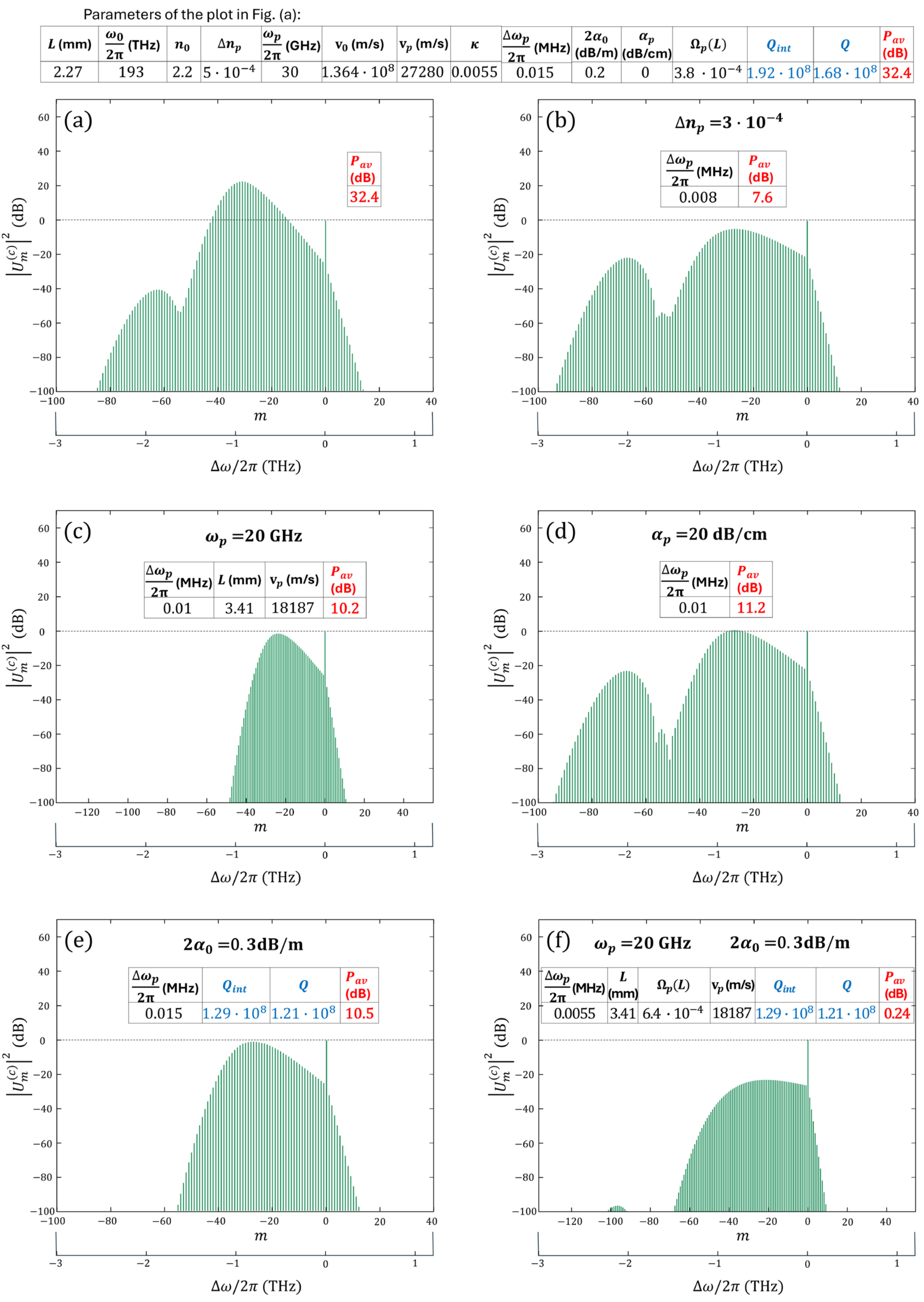
Parameters of the plot in Fig. (a):

| $L$ (mm) | $\frac{\omega_0}{2\pi}$ (THz) | $n_0$ | $\Delta n_p$ | $\frac{\omega_p}{2\pi}$ (GHz) | $v_0$ (m/s) | $v_p$ (m/s) | $\kappa$ | $\frac{\Delta\omega_p}{2\pi}$ (MHz) | $2\alpha_0$ (dB/m) | $\alpha_p$ (dB/cm) | $\Omega_p(L)$ | $Q_{int}$ | $Q$ | $P_{av}$ (dB) |
|---|---|---|---|---|---|---|---|---|---|---|---|---|---|---|
| 2.27 | 193 | 2.2 | $5 \cdot 10^{-4}$ | 30 | $1.364 \cdot 10^8$ | 27280 | 0.0055 | 0.015 | 0.2 | 0 | $3.8 \cdot 10^{-4}$ | $1.92 \cdot 10^8$ | $1.68 \cdot 10^8$ | 32.4 |

**Fig.10.** The resonant transmission power spectra for a racetrack resonator. The values of the system parameters of Fig. (a) are indicated in the table at the top of this figure. In plots (b)-(e), one parameter value from this table ($\Delta n_p, \omega_p, \alpha_p$, or $2\alpha_0$) is changed. The new value of this parameter is indicated at the top of each plot, followed by the table showing the new chosen modulation frequency offset $\Delta\omega_p$ and parameters whose values are changed accordingly. In plot (f), two parameters are changed ($\omega_p$ and $2\alpha_0$).

## B. The perturbation theory approach

The eikonal approximation used in Section IV.A does not allow us to consider sufficiently small values of $v_p$ since, according to Eq. (5), the condition of slowness of modulation in space restricts these values to $v_p \gg \omega_p v_0/\omega_0$. Remarkably, the latter restriction can be withdrawn for the parameters of our interest. Indeed, as shown in Section IV.A, the value of the modulation index $\Omega_p(L)$ required to arrive at the substantial effect of the amplification factor $F_m = \exp(-mG_p)$ is small, $|\Omega_p(L)| \ll 1$. Then, the solution of the wave equation, Eq. (1), can be found by the regular perturbation theory developed in Section III rather than by the eikonal approximation.

Here, we restrict our consideration to the forward wave propagation, assuming $k_p < k_0$, while the qualitatively different case including backward propagation and, in particular, Brillouin scattering, will be considered elsewhere [62]. Then, using the solution of the wave equation determined in Section III for modulation with attenuation (Fig. 8(b)), we determine the output amplitude $E_{out}(t)$ following Eqs. (46)-(48) where the expression for function $A(t)$ is now modified to

$$A(t) = \exp\left[ i\omega_0 T - \frac{\eta_0}{n_0}\omega_0 T + i\tilde{\Omega}_p(L)\cos\left(\omega_p t + i\tilde{G}_p(L)\right)\right]. \quad (57)$$

Here functions $\tilde{\Omega}_p(L)$ and $\tilde{G}(L)$ are defined by Eqs. (37)-(39). It is now straightforward to determine the output transmission amplitude $E_{out}(t)$ by comparing the expressions for $A(t)$ in Eqs. (49) and (57). It follows from this comparison that the comb spectral amplitudes $U_m^{(c)}$ of $E_{out}(t)$ can be found from Eq. (51) after the substitutions:

$$\Omega_p(L) \to \tilde{\Omega}_p(L), \quad iG_p \to i\tilde{G}_p(L). \quad (58)$$

Fig.10 shows the frequency comb spectra of a racetrack lithium niobate resonator with parameters indicated in the table at the top of this figure. Compared to Fig. 9, here we consider a much smaller phase velocity of the modulating wave, $v_p = 0.0002 v_0$, so that, for the considered modulation frequency $\omega_p = 2\pi \cdot 30$ GHz, we have $k_p = 0.772 k_0$. For this relation between the propagation constants, the eikonal approximation fails, while the optical propagation remains forward since $k_p < k_0$. The modulation frequency offsets $\Delta\omega_p$ in the plot of Fig.10(a) is chosen so that the bandwidth of the generated spectrum is $\Delta\omega_B/2\pi \cong 2.5$ THz. To optimize the amplification effect, we slightly modified the coupling coefficient to $\kappa = 0.0055$ compared to $\kappa = 0.005$ in Fig. 9. Then, as shown in Fig.10(a), for the modulation amplitude $\Delta n_p = 5 \cdot 10^{-4}$ (the same as in Fig. 9), we choose $\Delta\omega_p/2\pi = 0.015$ MHz to arrive at the total amplification of $P_{av} = 32.4$ dB. For a smaller modulation amplitude $\Delta n_p = 3 \cdot 10^{-4}$, we choose $\Delta\omega_p/2\pi = 0.01$ MHz to arrive at the total amplification of $P_{av} = 7.6$ dB (Fig.10(b)). Reduction of the modulation frequency to $\omega_p = 2\pi \cdot 20$ GHz and keeping the same relation between the propagation constants, $k_p = 0.772 k_0$, results in $P_{av} = 10.2$ dB (Fig.10(c)). In this case, the output comb bandwidth is reduced to $\Delta\omega_B/2\pi \cong 1$ THz. Figs.10(d) and (e) show that increasing the modulation wave amplitude attenuation from zero to $\alpha_p = 20$ dB/cm and the waveguide loss from $2\alpha_0 = 0.2$ dB/m (assumed in Fig. 9 and Fig.10(a)-(c)) to $2\alpha_0 = 0.3$ dB/m results in the reduction of amplification to $P_{av} = 11.2$ dB and $P_{av} = 10.5$ dB, respectively. Finally, Fig.10(f) shows that the amplification effect vanishes when the waveguide loss is below $2\alpha_0 = 0.3$ dB/m and, simultaneously, modulation frequency is below $\omega_p = 2\pi \cdot 20$ GHz. In the latter case, further decreasing the modulation frequency offset did not allow us to arrive at a significant amplification of light within the considered bandwidth $\Delta\omega_B/2\pi \sim 3$ THz.

## V. EXPERIMENTAL CHALLENGES

Current progress in the research and development of lithium niobate optical microresonators with exceptionally small losses [56, 57, 66] and eigenfrequency dispersion [67, 68], as well as in the design of microscopic RF electromagnetic and acoustic traveling wave generators [69-74], suggests that the system parameters required for the substantial amplification of light by electromagnetic and acoustic waves with dramatically smaller frequencies are potentially feasible. In this Section, we compare the microresonator and modulation parameters considered in Section IV with those experimentally achievable.

*Waveguide propagation loss.* The condition of amplification demonstrated in Section IV.B imposes a significant upper bound on the waveguide propagation loss of an optical resonator. This restriction can be relaxed by decreasing the ratio of phase velocities $v_p/v_0$ and increasing the modulation amplitude $\Delta n_p$. The phase velocity considered in Section IV.B is $v_p = 0.0002 v_0 \cong 27280$ m/s, while the modulation frequency and amplitude are $\omega_p = 2\pi \cdot 30$ GHz and $\Delta n_p = 5 \cdot 10^{-4}$. Practically, it is challenging to achieve so large modulation frequency and amplitude simultaneously. Here, these values were chosen to arrive at the smallest practically achievable waveguide loss of 0.2 dB/m required for the substantial amplification of light (see Fig.10(a)). A lithium niobate resonator with dramatically small waveguide loss 0.34 dB/m approaching the bulk material loss was demonstrated recently by chemo-mechanical waveguide polishing in Ref. [57]. The waveguide loss as small as 0.2 dB/m was demonstrated in Ref. [56] by post-fabrication annealing in oxygen atmosphere. Thus, the waveguides with losses required to experimentally realize the amplification with parameters of the transmission spectrum shown in Fig. 10 has been experimentally demonstrated.

*Resonator eigenfrequency dispersion*. The waveguide dispersion can be optimized to arrive at the smallest possible eigenfrequency dispersion. In contrast to the optimization for the optical frequency comb spectrum generated by optical microresonators commonly targeted at the largest possible bandwidth [28, 68], here we are interested in the accurate minimization of dispersion along a finite bandwidth $\Delta\omega_B$. In Section IV.B, we have $\Delta\omega_B \cong 2$ THz. The eigenfrequency dispersion of microresonators is characterized by the deviation from linear dependence $\delta\omega(\Delta\omega) = \omega_m - \omega_0 - m\omega_p$ of their spectral series $\omega_m$. Here $\Delta\omega$ is the continuous extrapolation of $m\omega_p$ and $m = \mathrm{int}(\Delta\omega/\omega_p)$ (see, e.g., [67, 68]). For the light frequency $\omega_0$ in the vicinity of the $\delta\omega(\Delta\omega)$ minimum, the amplification power will approach the values determined above in Sections IV.B if $\delta\omega(\Delta\omega_B)$ is much smaller than the resonance width $\Delta\omega_{res} = \omega_0/Q$ where the quality factor $Q$ is determined by Eq. (53), $|\delta\omega(\Delta\omega_B)| \ll \Delta\omega_{res}$. For the microresonator and modulation parameters leading to the amplification with the transmission spectra of Fig. 9, we have $Q{\sim}10^8$ and resonance width $\Delta\omega_{res}{\sim}2$ MHz. The value of deviation $\delta\omega(\Delta\omega_B)$ achieved in Ref. [67] for lithium niobate and in Ref. [68] for silicon nitride microresonators is smaller than 10 MHz over the bandwidth $\Delta\omega_B = 0.8$ THz and much smaller than 10 MHz over the bandwidth $\Delta\omega_B = 0.4$ THz. The dispersion required to realize the amplification effect described in Fig. 10 is an order of magnitude smaller. We suggest that microresonators with such a dispersion are potentially feasible.

*Modulation methods.* Different approaches have been developed for the effective optical waveguide modulation illustrated in Fig. 11 (see [19-28, 68-74] and references therein). The simplest design is represented by a spatially uniform capacitor, which can generate instantaneous modulation $\Delta n_p \cos(\omega_p t)$ corresponding to $v_p = \infty$ [37, 74] (Fig. 11(a)). A traveling wave refractive index modulation $\Delta n_0 \cos\left(\omega_p(t - x/v_p)\right)$ can also be introduced by an RF wave propagating parallel to the optical waveguide [75, 19-21] (Fig. 11(b)). This approach is beneficial for the modulation of photonic circuits with a traveling wave having the phase velocity $v_p$ comparable or equal to the phase velocity of light $v_0$. For a relatively small traveling wave phase velocity, $v_p \ll v_0$, the surface acoustic waves (SAWs) and bulk acoustic waves generated by an interdigital transducer (IDT) can be used [34, 71-73] (Fig. 11(c)). SAWs modulate the refractive index of an optical waveguide through the elasto-optics effect. An IDT tilted with respect to an optical waveguide by angle $\theta$ generates SAW propagating along the waveguide with the phase velocity $v_p = v_{sound}/\sin(\theta)$, where $v_{sound}$ is the speed of sound in the material. Thus, any traveling wave velocity $v_p$ exceeding $v_{sound}$ can be introduced. The characteristic refractive index variation induced by SAWs generated by an IDT in lithium niobate is estimated as $\Delta n_p \sim \frac{1}{2} n_0^3 (r_{33} - p_{33} d_{33}) V/w$, where $V$ is the voltage applied to the IDT, $w$ is the IDT finger separation, and we set the electro-optic coefficient $r_{33} = 30$ pm/V, the photoelastic coefficient $p_{33} = 0.1$, the piezoelectric coefficient $d_{33}$ =6 pm/V. Assuming $V/w \sim$1-10 V/μm we find $\Delta n_p \sim 10^{-4}$-$10^{-3}$ at the IDT position. Assuming that the EDT tilt angle $\theta$ and, thus, the SAW attenuation is small enough, we suggest that the modulation amplitude required for realization of amplification effect demonstrated in Fig.10 is potentially feasible.

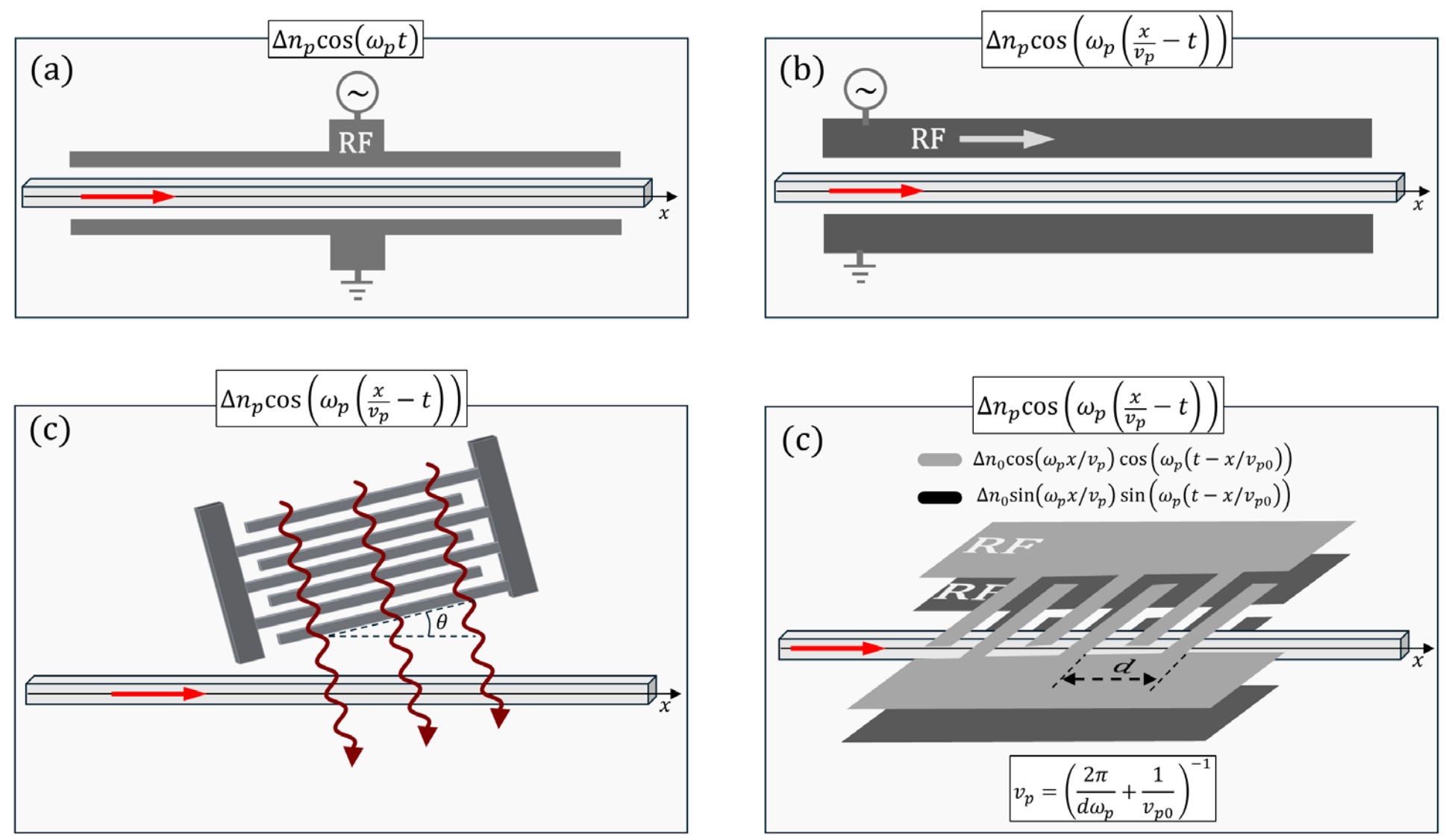

**Fig. 11.** Illustration of approaches to modulate the refractive index of an optical waveguide. (a) using a uniform capacitor; (b) using an RF waveguide; (c) using a tilted interdigital transducer (IDT); (d) using two phase-shifted IDT-shaped electrodes.

In a more general case, an advanced design of the modulators based on elasto-optic and electro-optic effects is required to arrive at the phase velocity, frequency, and spatial distribution of modulation. A single tilted IDT generating a traveling wave with velocity $v_p \ll v_0$ may be insufficient to generate the amplification of light due to the rapid SAW attenuation in space. However, a combination of in-phase tilted IDTs distributed along the optical waveguide and generating properly aligned SAWs may solve the problem. Alternatively, modulation of refractive index of an optical waveguide through a combination of bulk acoustic wave and Pockels electro-optic effects can be introduced by the IDT-type RF waveguides spatially modulated with the period $d$ and aligned along the optical waveguide (Fig. 2(d)). In this design, one RF waveguide introduces the optical waveguide's refractive index modulation equal to $\Delta n_0 \cos\left(\omega_p\left(t - x/v_{p0}\right)\right)\cos(2\pi x/d)$. In turn, another RF waveguide introduced modulation $\Delta n_0 \sin\left(\omega_p\left(t - x/v_{p0}\right)\right)\sin(2\pi x/d)$, which is phase shifted in time from the first one by $\pi/2$. The superposition of these modulations yields the traveling wave $\Delta n_0 \cos\left(\omega_p\left(x/v_p - t\right)\right)$ with the phase velocity $v_p = \left(\frac{2\pi}{d\omega_p} + \frac{1}{v_{p0}}\right)^{-1}$ which can be small for a small IDT period $d$. For example, for the case considered in Section IV.B and $v_{p0} \gtrsim v_0$, we have $d = 2\pi v_p/\omega_p \cong 1$ μm. Realization of such complex modulation structures is challenging since their proximity to the optical waveguide leading to the enhancement of modulation amplitude should be compromised with their effect on the optical waveguide loss.

*Suppression of nonlinear effects.* A critical requirement for maximizing the resonator Q-factor is the suppression of the nonlinear effects leading to the attenuation of light at the input frequency $\omega_0$. It has been shown in Ref. [57] that the experimentally observed Q-factor ~ $10^8$ in an LN ring resonator is not affected by nonlinear effects if the inter-resonator light power is ~ 1 mW or smaller. Then, for the cases considered in Fig.10 with coupling coefficient $\kappa = 0.0055$ and ~ 20 dB amplification, the output power will not exceed ~ 10 nW, while the input light power has to be as small as 0.1 nW.

## VI. DISCUSSION

We investigated the propagation of light with frequency $\omega_0$ through optical waveguides and racetrack resonators modulated by a traveling wave with relatively small frequency $\omega_p \ll \omega_0$ using the eikonal (semiclassical, WKB) approximation [9, 51, 52] (Sections II and IV.A) and a regular perturbation theory (Sections III and IV.B).

Section II of the paper was dedicated to the analysis of propagation of light in ideally dispersionless waveguides under the modulation of a traveling wave with a spatially uniform amplitude. We derived a general eikonal expression for the transmission amplitude similar to that derived several decades ago in application to the propagation of electromagnetic waves in transmission lines [8, 9]. Consequently, most of the results presented in this Section either resemble or complement the previously known results being now applied to the propagation of light. In particular, we showed that the effect of the traveling wave modulation can be significantly enhanced if modulation takes place along a sufficiently large waveguide length $L$ and in a close vicinity of the completely synchronous condition $v_p = v_0$. The determined transmission amplitude is periodic in time. However, its dependence on modulation length $L$ becomes aperiodic and growing for sufficiently small material losses if $|\mu| > 1$ and remains quasiperiodic if $|\mu| < 1$ [8-11]. We also showed that synchronous modulation is not advantageous compared to the commonly used instantaneous modulation for a relatively small modulation length $L$ of several tens of microns (see Figs. 4(a) and (b) and Figs. 5(c), (d) and (f)).

One of the major goals of this paper was understanding the feasibility of light amplification by a low frequency modulation of realistic optical waveguides and resonators. Having in mind realistic applications, we focused on the situations when light is propagating within a relatively small bandwidth $\Delta\omega_B(L) \ll \omega_0$. We found that, for a relatively large phase velocity of modulating wave, $v_p \gg v_0$, or for comparable velocities, $v_p \sim v_0$, the averaged over time amplification $P_{av}(L)$ is always small being proportional to $(\Delta\omega_B(L)/\omega_0)^2$ (see Eq. (31)). The situation becomes different for $v_p \ll v_0$ when Eq. (31) suggests that a moderate amplification of light may be possible within a small bandwidth $\Delta\omega_B(L)$. We found that such an amplification is currently very challenging to achieve experimentally in an open optical waveguide.

In Section IV we develop the theory of light propagating in a racetrack resonator modulated by a travelling wave with a relatively small frequency. Generally, a larger amplification can be achieved with a larger modulation frequency $\omega_p$ and amplitude $\Delta n_p$ and a smaller resonator waveguide loss $\eta_0$. To demonstrate the potential feasibility of amplification, we noticed that the transmission bandwidth $\Delta\omega_B$ of a resonator can be controlled and made small by increasing the offset of frequency $\omega_p$ from the exact modulation resonance. We found that, in the eikonal approximation, the amplification of individual comb lines increases with the parameter $-mv_0\omega_p/v_p\omega_0$, where $m$ is the frequency comp line number (see the expression for $G_p$ in Eqs. (51) and (27)). For the small modulation frequencies of our interest, $\omega_p \ll \omega_0$, this parameter increases with a decrease of the phase velocity of the modulating wave $v_p$. For very small $v_p \sim v_0\omega_p/\omega_0$, the eikonal approximation fails and in Section IV.B we apply the perturbation theory. Here we restrict the consideration by the case of forward propagation of light, while a qualitatively different case involving backward propagation and, in particular, Brillouin scattering, will be described elsewhere [62]. As an example, we demonstrated that the amplification of the full power of light as much as 30 dB can be achieved within the bandwidth $\Delta\omega_B \sim 2$ GHz in a racetrack resonator with the waveguide half-length $L =$ 2.27 mm and loss $2\alpha_0$ =0.2 dB/m modulated by an acoustic wave propagating along the waveguide with velocity $v_p = 0.0002v_0$=27280 m/s and modulation index $\Delta n_p = 5 \cdot 10^{-4}$ (Fig.10(a)). We showed in Section V that the required resonator waveguide and modulation parameters are potentially feasible.

While the developed theory assumes that the power of light inside the resonator is small enough not to introduce essential nonlinear effects, it is interesting to investigate if these effects can coexist with modulation-induced amplification. On the other hand, since the developed theory assumes very low input light power, it is interesting to explore its quantum version and investigate if the proposed device can amplify single-photon or few-photon signals.

## ACKNOWLEDGEMENTS

The author is grateful to A. A. Fotiadi, A. Gaeta, A. B. Matsko, D. V. Skryabin, J. Silver, S. K. Turitsyn, K. Vahala, and K. L. Vodopyanov for useful consultations and discussions of these results. The author extends his sincere gratitude to M. Lipson and her group for the detailed and fruitful discussion of the possible experimental realization of the effects described here and, in particular, to O. Florez, whose insightful question led to correcting the calculation mistake. This research was supported by the Engineering and Physical Sciences Research Council (Grants EP/W002868/1 and EP/X03772X/1), Leverhulme Trust (Grant RPG-2022-014), and Royal Society (Grant IES\R1\231250).

## DATA AVAILABILITY

The data that support the findings of this article are not publicly available. The data are available from the author upon reasonable request.

## Appendix A.

### Eikonal (WKB) approximation

We assume that the refractive index is a slow function of time and coordinates and formally introduce slow coordinate and time, $\zeta = \varepsilon x$ and $\tau = \varepsilon t$, where $\varepsilon \sim \omega_p/\omega_0 \ll 1$. We look for the solution of Eq. (1) in the form [50, 51]:

$$E(x,t) = \left(U_0(x,t) + \varepsilon U_1(x,t) + \ldots\right)\exp\left(\frac{i}{\varepsilon}S(x,t)\right). \qquad \text{(A1)}$$

Substituting Eq. (A1) into Eq. (1) and expanding the result in powers of $\varepsilon$, we arrive at a series of coupled equations for $U_m(x,t)$ and $S(x,t)$. In the zero order in $\varepsilon$, we obtain the equation for the eikonal $S(x,t)$ which determines the phase of solution:

$$n^2(x,t)S_t^2 - c^2 S_x^2 = 0\,. \qquad \text{(A2)}$$

This equation is reduced to the linear equation

$$n(x,t)S_t + cS_x = 0\,, \qquad \text{(A3)}$$

where it is assumed that the speed of light $c$ can have positive or negative sign. Once the solution $S(x,t)$ of Eq. (A3) is found, the amplitude terms $U_m(x,t)$ are determined from linear equations that can be solved successively. In particular, the zero-order term $U_0(x,t)$ of the amplitude of solution is found from the equation:

$$n(x,t)U_{0t}+cU_{0x}+\left(\frac{3}{2}n_t+\frac{c}{2n}n_x\right)U=0\,. \tag{A4}$$

For the case of the traveling wave refractive index defined by Eq. (2), solution of eikonal Eq. (A2) for the field phase and Eq. (A4) for the field amplitude can be found by the introduction of variables

$$\begin{aligned} t' &= t-\frac{x}{v_0}, \\ x' &= x, \end{aligned} \tag{A5}$$

where $v = c/n_0$ is the phase velocity of light (Fig. 1). Then, the refractive index in Eq. (2) depends on $t'$ only and Eqs. (A4) and (A5) can be rewritten down as

$$\left(a+b\cos(\omega_p t')\right)S_{t'}+v_0S_{x'}=0, \tag{A6}$$

$$\left(a+b\cos(\omega_p t')\right)U_{0t'}+v_0U_{0x'}-\delta(t')U=0, \tag{A7}$$

where

$$\begin{aligned} &a=1-\frac{v_0}{v_p}+i\frac{\eta_0}{n_0}, \quad b=\frac{\Delta n_p}{n_0}, \\ &\delta(t)=b\omega_p\sin(\omega_p t)\frac{2+b+3b\cos(\omega_p t)}{2\left(1+b\cos(\omega_p t)\right)}. \end{aligned} \tag{A8}$$

The general solutions of Eqs. (A3) and (A4) expressed through the original variables $x$ and $t$ are [76]

$$S(x,t)=\Phi_0\left(\xi(x,t)\right), \tag{A9}$$

$$\begin{aligned} &U_0(x,t)=\Phi_1\left(\xi(x,t)\right)W\left(t-\frac{x}{v_p}\right), \\ &W(t)=\exp\left(\int_0^t\frac{\delta(t)dt}{a+b\cos(\omega_p t)}\right)=\frac{(a+b)\sqrt{1+b}}{(a+b\cos(\omega_p t))\sqrt{1+b\cos(\omega_p t)}}. \end{aligned} \tag{A10}$$

Here $\Phi_k(\xi)$ are arbitrary functions determined by the boundary and initial conditions and

$$\begin{aligned} \xi(x,t) &= x-v\int_0^{t-\frac{x}{v_p}}\frac{dt}{a+b\cos(\omega_p t)} \\ &= x-\frac{2v}{\omega_p\sqrt{a^2-b^2}}\Xi\left(\sqrt{\frac{a-b}{a+b}},\left(\frac{\omega_p}{2}\left(t-\frac{x}{v_p}\right)\right)\right). \end{aligned} \tag{A11}$$

Here, function $\Xi(x,y)$ is the smooth continuation of $\arctan\left(x\cdot\tan(y)\right)$ as a function of $y$ which is convenient to calculate as

$$\Xi(x,y)=\int_0^y \frac{\partial}{\partial y}\left(\arctan\left(x,\tan(y)\right)\right)dy=\int_0^y \frac{dy}{x^2\sin^2(y)+\cos^2(y)} \tag{A12}$$

Ignoring the reflected wave, we determine the asymptotic solution of Eq. (1) corresponding to the boundary condition at $x = 0$ (Fig. 1)

$$E^{(in)}(x,t)=\exp\left[i\omega\left(\frac{x}{v_0}-t\right)\right],\quad x<0, \tag{A13}$$

separating it into the boundary conditions for $S(x,t)$ and $U_0(x,t)$:

$$S(0,t)=-\omega t\,, \tag{A14}$$

$$U_0(0,t)=1\,. \tag{A15}$$

Following the approach of Ref. [8], we introduce function $\bar{t}(\bar{\xi})$ inverse to function $\bar{\xi}(t) = \xi(t,0)$ which is found from Eq. (A11) as

$$\bar{t}(\bar{\xi})=-\frac{2}{\omega_p}\Xi\left(\sqrt{\frac{a+b}{a-b}},\frac{\omega_p}{2v_0}\sqrt{a^2-b^2}\,\bar{\xi}\right) \tag{A16}$$

where, again, function $\Xi(x,y)$ is the smooth continuation of $\arctan\left(x\cdot\tan(y)\right)$ as a function of $y$ defined by Eq. (A12). Using Eqs. (A11)-(A16), we find:

$$S(x,t)=-\omega\bar{t}(\xi(x,t))=\frac{2\omega}{\omega_p}\Xi\left(\sqrt{\frac{a+b}{a-b}},\frac{\omega_p}{2v_0}\sqrt{a^2-b^2}\,\xi\right) \tag{A17}$$

and

$$U_0(x,t)=\frac{W\left(t-\dfrac{x}{v_p}\right)}{W\left(\bar{t}(\xi(x,t))\right)}=\frac{\left(a+b\cos\left(\omega_p\bar{t}(\xi(x,t))\right)\right)\sqrt{1+b\cos\left(\omega_p\bar{t}(\xi(x,t))\right)}}{\left(a+b\cos\left(\omega_p\left(t-\dfrac{x}{v_p}\right)\right)\right)\sqrt{1+b\cos\left(\omega_p\left(t-\dfrac{x}{v_p}\right)\right)}}. \tag{A18}$$

**Appendix B.**

**The outgoing wave for the completely synchronous and lossless case $v_p = v_0$ and $\eta_0 = 0$**

In the synchronous lossless case, $v_p = v_0$ and $\eta_0 = 0$ (i.e., $a = 0$), Eqs. (A17) and (A18) are simplified. Setting $\Xi(x,y) = \arctan\left(x\cdot\tan(y)\right)$, we find

$$S_0(x,t)=\frac{2\omega}{\omega_p}\arctan\left(i\tan\left(\frac{i\omega_p bx}{2v_0}-\arctan\left(i\tan\left(\frac{\omega_p}{2}\left(t-\frac{x}{v_p}\right)\right)\right)\right)\right)$$

$$=\frac{2\omega}{\omega_p}\arctan\left(\frac{\tanh\left(\frac{b\omega_p x}{2v_0}\right)-\tan\left(\frac{\omega_p}{2}\left(t-\frac{x}{v_p}\right)\right)}{1-\tanh\left(\frac{b\omega_p x}{2v_0}\right)\tan\left(\frac{\omega_p}{2}\left(t-\frac{x}{v_p}\right)\right)}\right). \tag{B1}$$

Next, we simplify Eq. (A18) for the amplitude $U_0(x,t)$ under the same assumption $a=0$ assuming $\Delta n_p/n_0 \ll 1$. As the result we have:

$$U_0(x,t)=\frac{\cos\left(\omega_p\overline{t}(\xi(x,t)\right)}{\cos\left(\omega_p\left(t-\frac{x}{v_p}\right)\right)}. \tag{B2}$$

where

$$\cos\left(\omega_p\overline{t}(\xi(x,t)\right)=\left(\cosh\left(\frac{b\omega_p}{v_0}\xi(x,t)\right)\right)^{-1} \tag{B3}$$

Then, from Eqs. (B2) and (B3),

$$U_0(x,t)=\frac{1}{\cosh\left(\frac{b\omega_p x}{v_0}\right)-\sin\left(\omega_p\left(t-\frac{x}{v_p}\right)\right)\sinh\left(\frac{b\omega_p x}{v_0}\right)}. \tag{B4}$$

Averaging the output power, $P(x,t)=U_0(x,t)^2$, over time we find:

$$P_{av}(x)=\frac{\omega_p}{2\pi}\int_0^{2\pi/\omega_p}U_0(x,t)^2\,dt=\cosh\left(\frac{b\omega_p x}{v_0}\right). \tag{B6}$$

This equation coincides with Eq. (12) of the main text. It is seen from Eq. (B4) that for a sufficiently large modulation length $x$ corresponding to $\exp\left(\frac{b\omega_p}{v_0}\right)\gg 1$ the amplitude $U_0(x,t)$ rapidly changes with time in small intervals $\sin\left(\omega_p(t-x/v_p)\right)$ is close to unity. We determine the position $t_{max}$ of the maximum slope of $U_0(x,t)$ as a function of time by finding the zeros of its second derivative. Then, the condition of validity of the eikonal approximation, $\frac{\partial}{\partial t}U_0(x,t)\big|_{t=t_{max}}\ll\omega_0$ yields Eq. (13) of the main text.

**Appendix C.**

**The spectral bandwidth for the completely synchronous and lossless case $v_p=v$ and $\eta_0=0$**

We determine the transmission bandwidth by calculating the integral for the frequency comb amplitude given by Eq. (15) using the stationary phase method. For briefness, we introduce notations:

$$t_{xt} = \tan\left(\omega_p\left(\frac{x}{v_p} - t\right)\right), \quad t_x = \tanh\left(\frac{b\omega_p x}{2v_0}\right). \tag{C1}$$

The stationary phase time is then found by zeroing the derivative of the phase in the exponent of the integral of Eq. (15) where $E(x,t)$ is determined by Eq. (11):

$$\frac{\omega_0\left(t_x^2 - 1\right)\left(t_{xt}^2 + 1\right)}{t_x^2 t_{xt}^2 + t_x^2 + 4t_x t_{xt} + t_{xt}^2 + 1} + \omega_0 - \omega_p n = 0. \tag{C2}$$

From here, we find

$$t_{xt}^{\pm} = \frac{\left(\varsigma_+\varsigma_-\right)^{1/2}\left(t_x^2 - 1\right)^{1/2} \pm 2t_x\left(\frac{\omega_0}{\omega_p} - n\right)}{\left(n - 2\frac{\omega_0}{\omega_p}\right)t_x^2 - n}, \tag{C3}$$

$$\varsigma_{\pm} = n + \left(n \pm 2\frac{\omega_0}{\omega_p}\right)t_x. \tag{C4}$$

From Eq. (C3), the real stationary points exist only if $\Theta_+\Theta_- \geq 0$. After substitution of the expressions for $t_x$, $\varsigma_+$, and $\varsigma_-$ from Eqs. (C1) and (C4) into the latter inequality, we find the transmission band defined by Eq. (16).

**Appendix D.**

**Solution of the wave equation by the perturbation theory**

For sufficiently small modulation of the refractive index $\Delta n_p$ and small modulation index $|\Omega_p| \ll 1$, solution of the wave equation, Eq. (1), can be found by the perturbation theory. We rewrite Eq. (1) as

$$\left(\left(n_0 + \Delta n(x,t)\right)^2 E\right)_{tt} - c^2 E_{xx} = 0 \tag{D1}$$

and solve it by perturbations,

$$E(x,t) = E^{(0)}(x,t) + E^{(1)}(x,t) + E^{(2)}(x,t) \ldots, \tag{D2}$$

over $\Delta n(x,t)$. The general solution of Eq. (D1) is

$$E^{(gen)}(x,t) = E(x,t) + \Psi^+\left(t - \frac{x}{v_0}\right) + \Psi^-\left(t + \frac{x}{v_0}\right). \tag{D3}$$

where $\Psi^{\pm}(t)$ are arbitrary functions. The general solution of our interest, which is used in calculations of the transmission amplitude through an optical resonator corresponds to $\Psi^{-}(t) \equiv 0$, since it includes only optical waves propagating along a positive direction of axis $x$. In the zero, first, and second order, we have

$$n_0^2 E_{tt}^{(0)} - c^2 E_{xx}^{(0)} = 0\,, \tag{D4}$$

$$n_0^2 E_{tt}^{(1)} - c^2 E_{xx}^{(1)} = -2n_0 \left(\Delta n(x,t) E^{(0)}\right)_{tt}, \tag{D5}$$

$$n_0^2 E_{tt}^{(2)} - c^2 E_{xx}^{(2)} = -2n_0 \left(\Delta n^2(x,t) E^{(0)}\right)_{tt} - 2n_0 \left(\Delta n(x,t) E^{(1)}\right)_{tt}. \tag{D6}$$

Here we take into account the attenuation $\alpha_p$ of modulation along the waveguide length setting

$$\Delta n(x,t) = \Delta n_{p0} \exp\left(-\alpha_p x\right) \cos\left( \omega_p \left( t - \frac{x}{v_p} \right) \right). \tag{D7}$$

We choose the zero-order solution of Eq. (D1) as

$$E^{(0)}(x,t) = \exp\left[ i\omega_0 \left( \frac{x}{v_0} - t \right) \right]. \tag{D8}$$

To solve Eq. (D5) with $E^{(0)}(0,t)$ defined by Eq. (D8), we separate Eq. (D5) into two equations for $E^{(1)(+)}(x,t)$ and $E^{(1)(-)}(x,t)$:

$$n_0^2 E_{tt}^{(1)(\pm)} - c^2 E_{xx}^{(1)(\pm)} = -n_0 \Delta n_{p0} \left( \exp\left( \pm i\omega_p \left( \frac{x}{v_p} - t \right) + i\omega_0 \left( \frac{x}{v_0} - t \right) - \alpha_p x \right) \right)_{tt} \tag{D9}$$

Then, a particular solution of Eq. (D5) vanishing for $\Delta n(x,t) \to 0$ is

$$E^{(1)}(x,t) = E^{(1)(+)}(x,t) + E^{(1)(-)}(x,t)\,. \tag{D10}$$

Functions $E^{(1)(\pm)}(x,t)$ can be found in the form proportional to the right-hand side function of Eq. (D9):

$$E^{(1)(\pm)}(x,t) = \Delta U_0^{\pm} \exp\left( \pm i\omega_p \left( \frac{x}{v_p} - t \right) + i\omega_0 \left( \frac{x}{v_0} - t \right) - \alpha_p x \right). \tag{D11}$$

Substitution Eq. (D11) into Eq. (D9) yields:

$$\Delta U_0^{\pm} = \frac{\Delta n_p v_p^2 \left(\omega_0 \pm \omega_p\right)^2}{n_0 \left[ \pm\omega_p \left(v_0 - v_p\right) + i\alpha_p v_0 v_p \right] \left[ 2\omega_0 v_p \pm \omega_p \left(v_0 + v_p\right) + i\alpha_p v_0 v_p \right]}. \tag{D12}$$

It follows from Eq. (D11) that the perturbation component $E^{(1)(-)}(x,t)$ is a backward propagating wave if the wavenumber $k_p$ of the traveling wave is larger than the wavenumber $k_0$ of the input wave:

$$k_p > k_0,$$
$$k_p = \frac{\omega_p}{v_p}, \quad k_0 = \frac{\omega_0}{v_0} \tag{D13}$$
In this paper, we assume that the perturbed wave is forward propagating, i.e., that

$$k_p < k_0. \tag{D14}$$

Then, choosing an appropriate function $\Psi^{+}(t)$ in Eq. (D3), we find the first order solution of Eq. (D1) satisfying the boundary condition $E^{(1)}(0,t) = 0$:

$$E^{(1)}(x,t) = E^{(0)}(x,t)\sum_{\pm}\left(\Delta U_0^{\pm} \exp\left(\mp i\omega_p t\right)W^{\pm}(x)\right), \tag{D15}$$
$$W^{\pm}(x) = \exp\left(\pm i\frac{\omega_p x}{v_p} - \alpha_p x\right) - \exp\left(\pm i\frac{\omega_p x}{v_0}\right). \tag{D16}$$

This solution can be directly transformed to the form presented by Eqs. (36)-(39).